\documentclass[aps,prx,superscriptaddress,twocolumn, showpacs]{revtex4}
\usepackage{dcolumn}
\usepackage{graphicx} 
\usepackage{amsmath}
\usepackage{amsfonts}

\begin{document}

\title{Rayleigh anomalies and disorder-induced mixing of polarizations at nanoscale in amorphous solids. Testing 1-octyl-3-methylimidazolium chloride glass}

\author{M.G.~Izzo}
\affiliation{Universit\'a degli studi di Roma ''La Sapienza'', Dipartimento di Ingegneria Ingegneria Informatica Automatica e Gestionale Antonio Ruberti, Via Ariosto, 00185 Roma, Italy \\}
\affiliation{Istituto Italiano di Tecnologia - Center for Life Nanoscience, Viale Regina Elena, 291
00161 Roma, Italy \\}
\author{B. Wehinger}
\affiliation{University of Geneva, Department of Quantum Matter Physics, 24 Quai Ernest Ansermet, 1211 Genève 4, Switzerland}
\author{S.~Cazzato}
\affiliation{Universit\'a degli studi di Roma ''La Sapienza'', Dipartimento di Fisica, Piazzale Aldo Moro 5, 00185 Roma, Italy \\}
\affiliation{Chalmers University of Technology, Department of Applied Physics, Maskingr\"{a}nd 2, 412 58 Gothenburg, Sweden \\}
\author{A. Matic}
\affiliation{Chalmers University of Technology, Department of Applied Physics, Maskingr\"{a}nd 2, 412 58 Gothenburg, Sweden \\}
\author{C. Masciovecchio}
\affiliation{Sincrotrone Trieste S.C.p.A., S.S. 14 km 163,5 in AREA Science Park, I-34012 Basovizza, Italy \\}
\author{A. Gessini}
\affiliation{Sincrotrone Trieste S.C.p.A., S.S. 14 km 163,5 in AREA Science Park, I-34012 Basovizza, Italy \\}
\author{G.~Ruocco}
\affiliation{Universit\'a degli studi di Roma ''La Sapienza'', Dipartimento di Fisica, Piazzale Aldo Moro 5, 00185 Roma, Italy \\}
\affiliation{Istituto Italiano di Tecnologia - Center for Life Nanoscience, Viale Regina Elena, 291
00161 Roma, Italy \\}

\date{\today}

\begin{abstract}
Acoustic excitations in topologically disordered media at mesoscale present anomalous features with respect to the Debye's theory. In a three-dimensional medium an acoustic excitation is characterized by its phase velocity, intensity and polarization. The so-called Rayleigh anomalies, which manifest in attenuation and retardation of the acoustic excitations, affect the first two properties. The topological disorder is, however, expected to influence also the third one. Acoustic excitations with a well-defined polarization in the continuum limit present indeed a so-called mixing of polarizations at nanoscale, as attested by experimental observations and Molecular Dynamics simulations. We provide a comprehensive experimental characterization of acoustic dynamics properties of a selected glass, 1-octyl-3-methylimidazolium chloride glass, whose heterogeneous structure at nanoscale is well-assessed. Distinctive features, which can be related to the occurrence of the Rayleigh anomalies and of the mixing of polarizations are observed. We develop, in the framework of the Random Media Theory, an analytical model that allows a quantitative description of all the Rayleigh anomalies and the mixing of polarizations. Contrast between theoretical and experimental features for the selected glass reveals an excellent agreement. 
The quantitative theoretical approach permits thus to demonstrate how the mixing of polarizations generates distinctive feature in the dynamic structure factor of glasses and to unambiguously identify them. The robustness of the proposed theoretical approach is validated by its ability to describe as well transverse acoustic dynamics.    
\end{abstract}

\maketitle
The modelling of acoustic dynamics properties in amorphous solids at mesoscale is strongly linked to the understanding of the origin of their  macroscopic anomalies, such as the hump in the specific heat of glasses at about $10$ $K$ \cite{Phillips} and excess over the Debye level of the Vibrational Density of States (VDOS) at energies of few $meV$, called Boson Peak (BP) \cite{Shintani,Shir_gen1,Schober1,Marruzzo, Chumakov1, Chumakov2}. Though an unanimous theory was not established, it is worth to notice that all the developed theories deal with the understanding of how elastic disorder affects the acoustic excitations at nanoscale, whether the disorder is treated as only a small perturbation with respect to the ordered structure of the crystals \cite{Chumakov1,Giordano}, modelled as defects embedded in an otherwise homogeneous medium \cite{Schober1,Sheng} or represented by spatial fluctuations of elastic moduli - local \cite{Schirmacher_4th,Schi_SCBorn,Kohler,Shir_gen1,Marruzzo,Ferrante} or with long-range correlations \cite{Tanaka}. 
Disorder on mesoscopic scale generates an inhomogeneous spatial distribution of the elastic constants \cite{Schirmacher_4th,Schi_SCBorn,Kohler,Shir_gen1,Schober1,Marruzzo,Ferrante,Sheng, Tanaka}. 
In a statistical framework the local spatial distribution of elastic constants can be described by introducing a spatial correlation function of their fluctuations with respect to the average value on the whole system's volume \cite{Schirmacher_4th,Schi_SCBorn,Kohler,Shir_gen1,Marruzzo,Ferrante,Sheng, Tanaka}. The related correlation length corresponds to the average value of the radius of spatial domains where the local elastic constants remains roughly constant \cite{Schirmacher_4th,Schi_SCBorn,Kohler,Shir_gen1,Sheng,Torquato, Calvet, Turner}. We will refer in the following to them as to inhomogeneity or heterogeneity domains.    

Acoustic excitations of glasses at mesoscale are affected by the so-called Rayleigh anomalies \cite{Schirmacher_4th,Schi_SCBorn,Kohler,Shir_gen1,Schober1,Marruzzo,Ferrante, Tanaka, Shintani, Ruffle2, hydro_MonGio, Mossa}. They consist in a strong increase of the attenuation and a corresponding softening of the phase velocity. 
An acoustic wave traveling in a three-dimensional medium is characterized by its phase velocity, amplitude and \textit{polarization}.  The presence of topological disorder is thus expected to affect \textit{all} these properties. The Rayleigh anomalies highlight how disorder can affect the first two. Concerning polarization, elementary elasticity theory states that a purely longitudinal (or transverse) wave impinging on an interface between two different elastic media is transformed in waves with mixed polarization \cite{Landau}. Acoustic excitations with mixed polarization at nanoscale have been indeed observed by both Inelastic X-ray/Neutron Scattering (IXS/INS) experiments \cite{Ruzi, Scopigno1HFglasses_transverse, zanatta_INS_GeO2,Cimatoribus, Cunsolo, Benci, Bolmatov}, as well as by Molecular Dynamics (MD) simulations \cite{Sampoli, Bryk1,Bryk2,Ribeiro2} in several amorphous solids and liquids in the first pseudo-Brillouin zone, typically at wavevectors of magnitude of $4-5$ $nm^{-1}$. Despite this, one meets the lacking of an analytical model addressing the occurence of disorder-induced mixing of polarizations. The mixing of polarizations, furthermore, has never been related to the Rayleigh anomalies by a quantitative theory.  We propose to cope with these shortcomings. 

We provide in the present manuscript a detailed experimental characterizations of vibrational dynamics properties in 1-octyl-3-methylimidazolium chloride, [C8MIM]Cl, glass. The wavevector-resolved acoustic dynamics features have been probed by IXS. In particular, the acoustic dynamics of a glass in the wavevectors range where the mixing of polarizations onsets has been characterized in this work with unprecedented precision. Proper INS measurements yield the corresponding VDOS. Distinctive features which can possibly be related respectively to Rayleigh anomalies, mixing of polarization and presence of the BP are highlighted by a phenomenological data analysis, in line with a general approach largely exploited and attested in literature \cite{Giordano, hydro_MonGio, Ruffle2, Ruzi, Scopigno1HFglasses_transverse, Cimatoribus, Benci, Cunsolo}. Such a features are discussed in details in Sec. \ref{exp_results}. They are in agreement with literature-reported experimental and MD results obtained for many different amorphous solids \cite{Giordano,Ferrante, hydro_MonGio, Ruffle2, Ruzi, Scopigno1HFglasses_transverse, zanatta_INS_GeO2, Cimatoribus, Benci, Cunsolo, Bolmatov}. The choice of the selected glass in order to address the above-mentioned matter rely on the fact that its structural heterogeneity at nanoscopic scale has been largely attested and characterized by Small Angle X-ray or Neutron Scattering measurements and by MD simulations \cite{Fujii,Bodo,Margalius, Ferde, Aoun, Hardacre}. The value of the average radius of heterogeneity domains, $a$, can be obtained. Since the mixing of polarizations is expected to onsets when the magnitude of wavevector, $q$, becomes comparable to $a^{-1}$ \cite{Calvet}, its exact knoweldge permits to relate  experimental features emerging at such $q$'s, highlighted by the phenomenological data analysis, to the mixing of polarizations. More importantly, as discussed in the following, the knoweldge of the value of $a$ permits to fix non-trivial input parameter of the theoretical approach we developed, thus making the experimental observations on the selected glass a reliable base for a genuine testing of the theory.  
The value of $a$, in particular, can be derived from the wavevector position, $q_{FSDP}$, of the First Sharp Diffraction Peak (FSDP) in the static structure factor, $S(q)$, as discussed in Sec. \ref{exp_results}. 
Among the large class of Ionic Liquids, [C8MIM]Cl has been selected because 1) the wavevector location of the FSDP matches the wavevectors region typically probed by IXS; 2) the width of the FSDP is sufficiently small to permit to safely neglect the distribution of the size of the heterogeneity domains around their average value, i.e. the value of $a$ is well defined. 

Modelling the amorphous solid as a random medium characterized by local elastic constants fluctuating in space, provides a basis for the definition of a stochastic Helmholtz equation and of a related Dyson equation, introduced in the framework of the Random Media Theory (RMT) \cite{Calvet, Turner, B1,Sobczyk, Marruzzo, Ferrante, Schirmacher_4th,Schi_SCBorn,Kohler,Shir_gen1}. The Dyson equation describes the ensemble averaged elastodynamic response of the system to an impulsive force. This formalism can thus in principle represents a valuable tool to achieve a full description of acoustic excitations properties in glasses, including the Rayleigh anomalies and the mixing of polarizations. The Dyson equation, however, can be solved only by recourse to suitable approximations, which hold in a limited $q$, or frequency, $\omega$, range \cite{Sobczyk, Calvet, Schirmacher_4th, Marruzzo, Ferrante, Kohler,Schi_SCBorn,B1,Turner, Bourret}. This can possibly avoid the attainment of the purpose because the above cited phenomena arise in different $q$ regions.
The Rayleigh anomalies appear when $q$ is  much lower than the inverse of $a$, typically of the order of $nm$ \cite{Marruzzo, Ferrante, Calvet, Schirmacher_4th}, whereas  the coupling between longitudinal and transverse excitations is maximum when $q$ becomes comparable to $a^{-1}$ \cite{Calvet}. Approximate solutions of the Dyson equation revealed to catch the characteristic behavior of acoustic excitations in glasses in the Rayleigh region \cite{Marruzzo, Ferrante, Schirmacher_4th} but RMT approach has not yet been used to describe depolarization effects of acoustic excitations in glasses.
The statistical description of spatial distribution of elastic constants makes results obtained in a RMT framework dependent on some input parameters, such as the value of $a$ or the strength of spatial fluctuation \cite{Ferrante, Marruzzo, Calvet, Turner, Bourret, B1, Sobczyk}. Only in rare cases they can be independently determined by proper experiments, being rather chosen, usually, as \textit{ad hoc} parameters in order to reproduce the observed characteristics of the acoustic excitations. This can give rise to tautology in the experimental corroboration of the theory. In particular, the knowledge of the exact value of $a$ becomes all the more relevant for the purpose to theoretically describe phenomena, such as the mixing of polarizations, occurring when $q$ matches the inverse of $a$. The choice of the proper testing material becomes thus an integral part of theoretical developments. 
The comparison between experimental and theoretical features provided in this work is excellent, allowing the attribution of specific features observed in the dynamic structure factor of glasses to the mixing of polarizations.    

While in Ref. \cite{Izzo2} it is stressed how the proposed theoretical model, the so-called Generalized Born Approximation (GBA), can quantitatively describe all the Rayleigh anomalies and the mixing of polarizations at mesoscale in glasses, in the present manuscript we further provide details on the phenomenological data analysis of acoustic dynamics properties of [C8MIM]Cl glass and compare them with general insights for amorphous solids found in literature \cite{Giordano,Ferrante, hydro_MonGio, Ruffle2, Ruzi, Scopigno1HFglasses_transverse, zanatta_INS_GeO2, Cimatoribus, Benci, Cunsolo, Bolmatov}. Moreover, the framing of GBA in the RMT is discussed, as well as the comparison with previous approaches developed in the same framework, better clarifying its physical ground and why a generalization of the so-called Born approximation is required. The features of acoustic transverse dynamics on the mesoscopic region derived from GBA are furthermore analysed.
Description of the experiments and specifics of the GBA calculations are reported with a detail sufficient to permit the reproduction of both experimental and theoretical results. 

This paper is organised as follows. Sec. \ref{exp_results} contains a description of experiments, of the phenomenological data analysis and discusses the experimental results. Sec. \ref{GBA} introduces to the GBA, discusses its framing in the RMT and explains how theoretical results are produced. Sec \ref{discussion} shows the general understanding of longitudinal and transverse acoustic dynamics of amorphous solids on mesoscopic region provided by GBA, while discussing how it can quantitatively reproduce experimental results for [C8MIM]Cl glass. Concluding remarks are presented in Sec. \ref{conclusion}. Appendix A contains specifics of GBA calculations. Appendix B discusses the coupling of acoustic excitations, modelled by GBA, with  Intermolecular Vibrational Modes (IVMs). 
\section{Results}
\subsection{Experimental characterisation of longitudinal acoustic dynamics and Vibrational Density of States in 1-Octyl-3-methylimidazolium chloride ionic glass} \label{exp_results}
\paragraph{The Inelastic X-ray Scattering experiment.} \label{IXS_Methods} The IXS experiment was carried out at the ID28 beamline of the European Synchrotron Radiation Facility (ESRF). The data were collected in the $q$ range $[1-15]$ $nm^{-1}$ with a $q$ step of about 0.4 $nm^{-1}$. The experiment was performed at $23.725$ keV, using the Silicon $(12,12,12)$ reflection, which provides an overall energy resolution of about $1.4$ meV (FWHM), determined from a plexiglass slab at T=10 K. The $q$-resolution was set to 0.18 $nm^{-1}$ for the first $q$ points (i.e. 1.0 $nm^{-1}$, 1.4 $nm^{-1}$, 1.8 $nm^{-1}$) and fixed to 0.37 $nm^{-
1}$ for the remaining $q$'s. The energy spectra were collected in the $-30$ to $30$ meV range. The
sample cell was an Al tube of $5$ mm length capped with two oriented monocrystal diamond windows ($0.5$ $mm$ thick). The [C8MIM]Cl sample was loaded in the sample cell in an Ar-filled glove box to avoid water
contamination. A cryostat was used to cool the sample down to the glassy state. The background contribution from the sample environment was measured in the same experimental configuration as the sample and subtracted from the data, thus obtaining the sample signal, $I(q,E)$. This latter provides a measure of the dynamic structure factor related to longitudinal acoustic dynamics, $S_L(q,E)$, via the expression, $I(q,E)=A(q) [E \frac{n(E)+1}{k_BT} S_L(q,E)] \otimes R(E)$, where $E=\hbar \omega$ is the exchanged energy between the probe and the sample, $\hbar$ is the reduced Planck constant, $n(E)$ is the Bose factor, $R(E)$ is the experimental energy resolution and $\otimes$ represents the convolution operation. $A(q)$ is an overall intensity factor.
\paragraph{The Inelastic Neutron Scattering experiment.}
The INS experiment was performed with the MARI spectrometer at the ISIS spallation neutron source in Rutherford Appleton Laboratory. The sample was loaded in an Al made cylindrical annular can with a thickness corresponding to a sample transmission $T=0.9$. The incident neutron energy was 15 $meV$. The temperature of the sample was tuned by a cryostat. The empty cell scattering, $I_{EC}(q,E)$, was measured under the same sample experimental conditions. The sample signal, $I(q,E)$, is obtained from the total signal, $I_t(q,E)$, through the relationship $I(q,E)=I_t(q,E)-TI_{EC}(q,E)$. Multiple scattering and self-absorption are assumed to be negligeable for the given sample geometry and transmission coefficient. For [C8MIM]Cl the incoherent neutron scattering cross section, $\sigma_{incoh}$, is much larger than the coherent one, $\sigma_{coh}$, being $\frac{\sigma_{coh}}{\sigma_{incoh}}=0.07$.
The measured data  are integrated over $q$ in the largest range available from the experimental configuration, i.e. $20<q<40$ $nm^{-1}$. It is thus obtained the averaged VDOS, neutron-weighted over the atomic species. The reduced VDOS, $\frac{g(E)}{E^2}$, is obtained from the integrated signal, $I(E)$, through the relationship
$\frac{g(E)}{E^2}=A \frac{I(E)}{E[n(E)+1]}$, where $A$ is an intensity factor. 

\medskip

\begin{figure}
\centering
\includegraphics[width=1.0\linewidth]{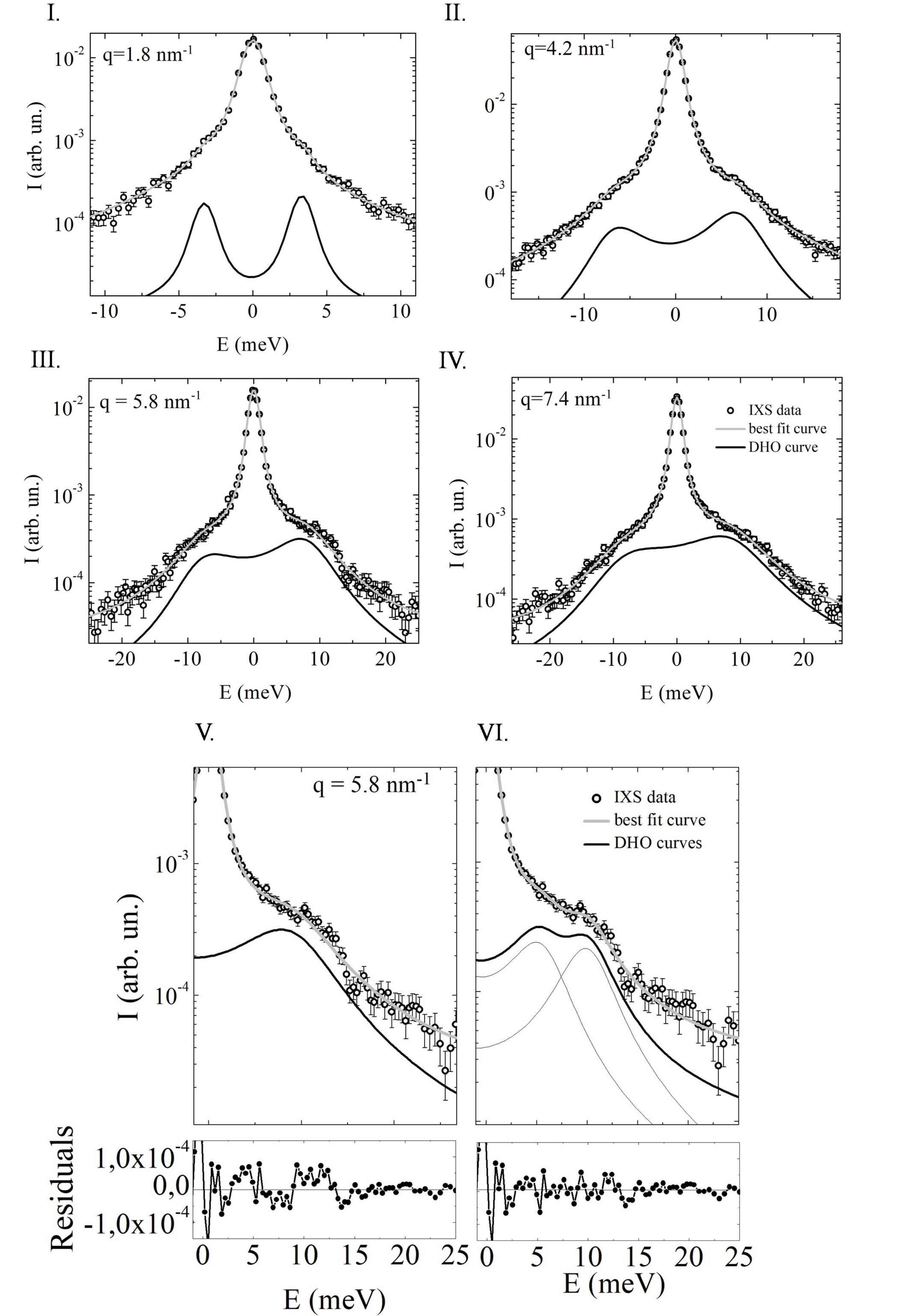}
\caption{\textit{Panels I-IV.} IXS spectra of $[C8MIM]Cl$ glass at $T= 176$ $K$ for selected $q$-values (black circles with error bars), best-fit curves obtained by 1-DHO fitting model (grey lines) and inelastic components of best fit curve (black lines). \textit{Panels V-VI.} Detail of a representative IXS spectrum (black circles with error bars) in the high-$q$ region. Best fit curves (grey lines) and inelastic contributions (black lines) obtained by respectively using 1-DHO fitting model and 2-DHO fitting model are shown. Corresponding fit residuals are displayed in the bottom panels.}
\label{esempio_spettri}
\end{figure}
\begin{figure}
\centering
\includegraphics[width=0.8\linewidth]{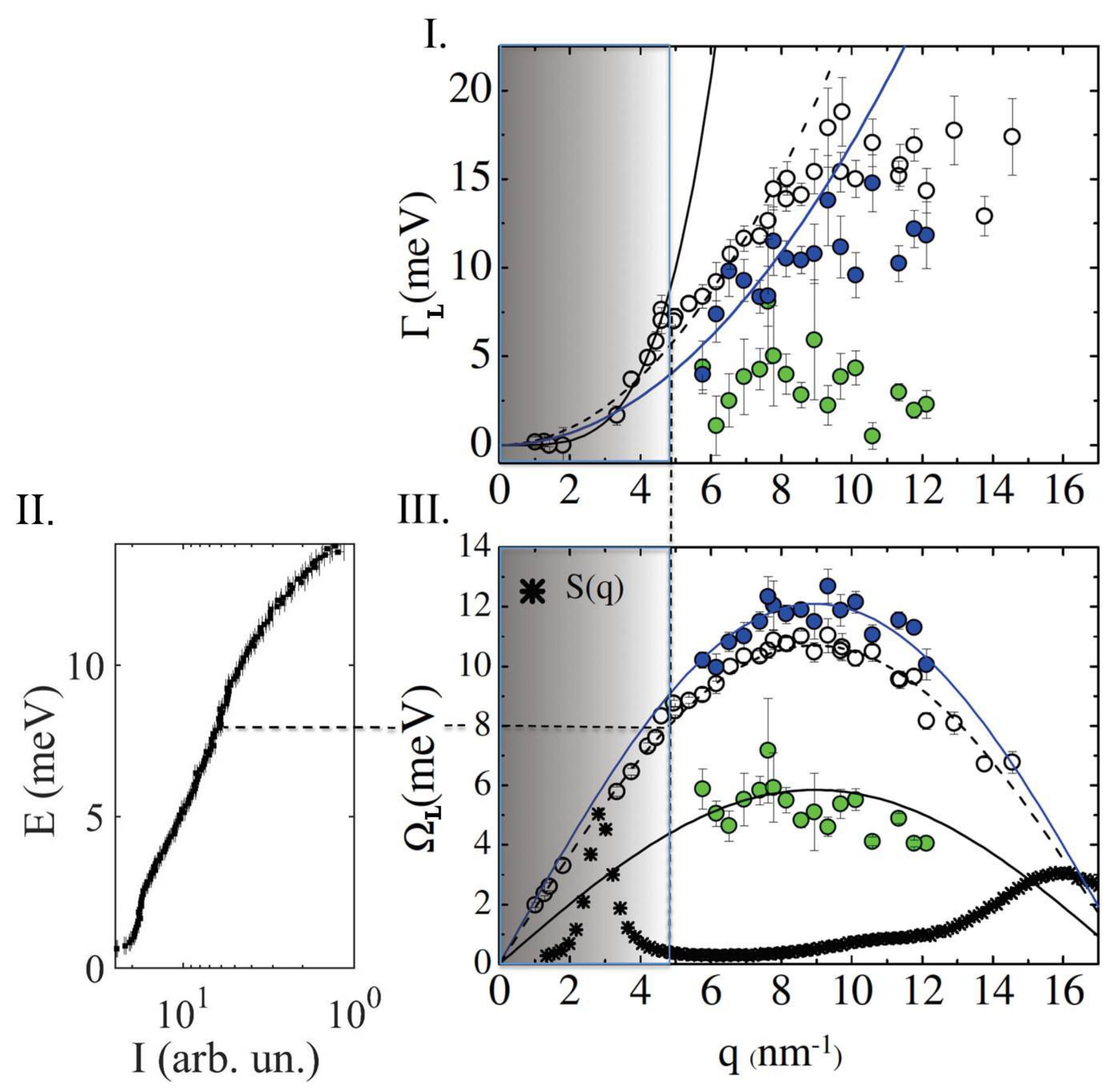}
\caption{\textit{Panel I.} Broadening ($\Gamma_L$) as a function of $q$ derived from the fitting of the IXS data by 1-DHO fitting model (open circle) and 2-DHO fitting model in the  high-$q$ region (blue and green circles). Full black line shows the $q^4$ trend reproducing the $\Gamma_L$ dispersion in the low-q region. Blue and dashed black lines are guide to eye displaying the $q^2$ trend of the broadenings in the high-$q$ region. \textit{Panel II.} Reduced VDOS, $g(E)/E^2$, obtained by INS measurements. \textit{Panel III.} Characteristic energy ($\Omega_L$) of the inelastic excitations as a function of $q$, obtained by fitting with 1-DHO fitting model (open circle) and 2-DHO fitting model in the high-$q$ region (blue and green circles). The dashed, blue and black lines reproduce the dispersion curves of the respective $\Omega_Ls$. They are obtained by the sinusoidal function, $c  \  \frac{q_0}{\pi}sin(\frac{q\pi}{q_0})$ \cite{Bosak}, where $c$ and $q_0$ are adjustable parameters. The X-ray-weighted static structure factor is shown (stars).} \label{dispersion1DHO}
\end{figure}

Experimental characterisation of longitudinal acoustic dynamics and VDOS of [C8MIM]Cl at $T=176$ $K$ ($T_g = 214$ $K$) was achieved respectively by IXS and INS. Fig. \ref{esempio_spettri}  shows IXS spectra for selected $q$ values. High quality data were obtained at all wavevectors and an inelastic contribution is detectable in all the spectra, except for those corresponding to $q$ close to $q_{FSDP}$, where the spectrum is dominated by the strong elastic signal. $S_L(q,E)$ can be modelled with a sum of a delta function for the elastic component and a Damped Harmonic Oscillator (DHO) function for the inelastic component (1-DHO fitting model).  Such a protocol provides a measure of the characteristic energy and broadening (attenuation) of the longitudinal acoustic excitation for each $q$, respectively $\Omega_L(q)$ and $\Gamma_L(q)$.  The values of $\Omega_L(q)$ and $\Gamma_L(q)$ with the corresponding error bars obtained by exploiting the 1-DHO fitting model are displayed in Fig. \ref{dispersion1DHO}, \textit{Panels I} and \textit{III}, by open circles. The high quality of the IXS data acquired permits to assign a value to $\Omega_L$ and $\Gamma_L$ even at small values of $q$, in the interval $1-2$ $nm^{-1}$.
In the high-$q$ region ($q>5$ $nm^{-1}$) an additional feature in the inelastic wings of the IXS spectrum is observed, in agreement with experimental observation in several glasses, see e.g. Refs. \cite{Ruzi,Scopigno1HFglasses_transverse,zanatta_INS_GeO2}. Similar features are commonly related to \lq projection' of transverse into longitudinal dynamics. To empirically account for the presence of this extra feature, the inelastic part of the spectrum can be fitted with a model function that includes two DHO functions (2-DHO fitting model). The $S_L(q,\omega)$ is thus empirically modelled by the expression
\begin{eqnarray}
& S_L(q,E)=\sum_{n} \frac{I_{(n)}\Gamma_{L(n)}\Omega_{L(n)}^2}{[{\Omega_{L(n)}}^2-E^2]^2+E^2 {\Gamma_{L(n)}}^2}+ I_0 \delta(E), \ \
\end{eqnarray}
where $n=1$ for the 1-DHO fitting model and $n=1,2$ for the 2-DHO fitting model. Fig. \ref{esempio_spettri} shows a magnification of the inelastic component of an IXS spectrum acquired at $q>5$ $nm^{-1}$. Best fit curves and fit residuals corresponding to 1-DHO and 2-DHO fitting models are included for comparison. Inspection of fit residuals corresponding to 1-DHO fitting model shows the presence of structured features revealing indeed inadequacy in the choice of this fitting model function. Fig. \ref{dispersion1DHO}, \textit{Panels I} and \textit{III}, show the best fit values of $\Omega_L$ and $\Gamma_L$ corresponding to each one of the two DHO functions composing the fitting model function for $q>5 \ \ nm^{-1}$, displayed respectively by blue and green full circles. 
We will analyse both best fit results obtained by fitting with the 1-DHO fitting model in the whole measured $q$-region and with the 2-DHO fitting model in the only high-$q$ region. The trend of $\Omega_L(q)$ and $\Gamma_L(q)$ obtained by the 1-DHO fitting model presents at $q \sim a^{-1}$ features that can be related to the mixing of polarizations, described in points (ii) and (iii) in the following and examined in Sec. \ref{discussion}. 
As emphasized in Sec. \ref{discussion} the experimental spectrum at each measured $q$ can be fully reproduced by GBA based calculations. This makes the fitting with 1-DHO or 2-DHO model functions an experimentally convenient, yet theoretically unnecessary, phenomenological expedient.
%

%
Fig. \ref{dispersion1DHO} shows $g(E)/E^2$ (\textit{Panel II}) and $S(q)$ (\textit{Panel III}), measured by X-ray scattering. The reduced VDOS presents 1) a peak around $2$ $meV$, referred to as the Boson Peak (BP) \cite{Ribeiro_Raman2}; 2) a broad peak-like feature at higher energies, in the region between $7$ and $12$ $meV$, related to librational modes of the imidazolium ring \cite{Ribeiro_Raman2} - referred hereafter IVMs. 
The static structure factor shows a FSDP at $q_{FSDP} = 2.8$ $nm^{-1}$, which is related to nanoscale segregation of the cations alkyl chains \cite{Fujii,Ferde}. Upon definition of the ring centres and anions as polar entities and the alkyl chains as nonpolar ones, the segregation of the alkyl chains results in a local structure formed by an alternation of polar and nonpolar domains \cite{Aoun}. As proved by Neutron and X-ray scattering experiments \cite{Margalius, Hardacre, Fujii}, the FSDP is related to the scattering signal coming from those anions that are on the side of the charged imidazolium ring belonging to cations that have opposite alkyl tails (see the pictorial representation in Fig. \ref{fig_Sq}).
The characteristic length scale, $2 \pi/q_{FSDP}$, gives thus an estimation of the diameter of nonpolar domains \cite{Fujii}, i.e. $2a \sim 2 \pi /q_{FSDP}$. Since, furthermore, intermolecular forces acting in polar and apolar regions are of different nature, it is possible to assume that \textit{a)} to the heterogeneous local structure it corresponds a heterogeneous spatial distribution of the local elastic constants \cite{Ribeiro2}; \textit{b)} the difference in the elastic constants between the two kind of domains, which corresponds to the intensity of elastic fluctuations, is quite large.
Inspection of Fig. \ref{dispersion1DHO} permits to conclude that: (i) in the low-$q$ region (grey-shaded region in Fig. \ref{dispersion1DHO}, \textit{Panels I} and \textit{III}) the Rayleigh anomalies are present. The increase of the acoustic attenuation that roughly follows a $q^4$ trend is clearly observable. The softening of the phase velocity is better highlighted in Fig. \ref{MODEL_disp2_ord2}, but it can be perceived, even if with lower clarity, also in the $q$ trend of $\Omega_L$s displayed in Fig. \ref{dispersion1DHO}. (ii)  It exists a crossover in the wavevector dispersion of $\Gamma_L$ at $q_c=4.8 \  nm^{-1}$ from a $q^4$ to a $q^2$ trend. We observe that $q_{c} \sim 2q_{FSDP} \sim 2 \pi /a$, i.e. the crossover appears at wavevectors related to the typical size of elastic heterogeneity domains. (iii) There is a kink at $q_c$ in the $\Omega_L$ dispersion. (iv) For $q>q_c$ the inelastic wings of $S_L(q,E)$ are composed by two features.
The crossover in the $\Gamma_L$ trend has been observed in several other glasses, see e.g. Refs. \cite{hydro_MonGio, Ruffle2}. It usually occurs at frequencies in the region of the BP, at slightly lower frequency than in the present case.
We show in Sec. \ref{discussion} how the behaviour described in points (i)-(iv) can be quantitatively described by the proposed stochastic approach derived in the framework of the RMT, which accounts for the elastic heterogeneity of the disordered system. In particular the features observed at $q$ of the order of or larger than $q_c$ can be related to the mixing of polarizations. We notice, however, that the value of $\Omega_L(q_c)$ ($\sim 8 \ meV$) belongs to the energy range of the VDOS where it is observed the broad feature related to IVMs, which can thus be brought into play in the description of the acoustic excitations behaviour  at such wavevectors. The possible coupling between acoustic excitations and IVMs is discussed in Appendix B.
\subsection{Generalized Born Approximation and assesment of theoretical outputs} \label{GBA}
\paragraph{Random Media Theory and Generalized Born Approximation.} Purpose of the RMT is to describe the space (\textbf{r}) and time (t) evolution of the ensemble averaged response of a spatial heterogeneous medium after the application of an external input in the space-point \textbf{r'}. The system's response is formally expressed by the second-rank averaged Green's dyadic $<\textbf{G}(\textbf{r},\textbf{r'},t)>$, solution of the corresponding Dyson equation. The brackets $< \ \ >$ denotes ensemble average. The bold font style displays a tensor. The disordered heterogeneous spatial distribution of fluctuating quantities is statistically described by their space-correlation function. By virtue of the fluctuation-dissipation theorem, the dynamic structure factors related to both longitudinal and transverse dynamics, can be expressed as a function of the averaged Green's function, $<G_{L(T)}(\textbf{q},\omega)>$, i.e. $S_{L(T)}(\textbf{q},\omega) = \frac{1}{\pi} \frac{q^2}{\omega} Im\{<G_L(\textbf{q},\omega)>\}$. The wavevector, $\textbf{q}$, and $\omega$, are the conjugate Fourier variables of $\textbf{r}$ and $t$, respectively.
We let the only elastic tensor, $\textbf{C}(\textbf{r})$, to fluctuate in space. A formal statement of the Dyson equation provides for the introduction of a self-energy or mass-operator, $\boldsymbol{\Sigma}(\textbf{q},\omega)$, which embeds all the information related to system's inhomogeneity,
\begin{eqnarray}
<\textbf{G}(\textbf{q},\omega)>= [\textbf{G}^0(\textbf{q},\omega)^{-1}-\boldsymbol{\Sigma} (\textbf{q},\omega)]^{-1} \label{Dyson}.
\end{eqnarray}
where $\textbf{G}^0(\textbf{q},\omega)$ is the bare Green's dyadic describing the \lq bare' system in the absence of spatial fluctuations. 
The self-energy can be formally cast by a perturbative Neuman-Liouville series expansion \cite{Sobczyk, B1}. Through the introduction of the Feynman diagram technique it is possible to establish a correspondence between the terms composing the perturbative series expansion and the scattering events of the elastic perturbation in the medium \cite{B1}.
Under the hypothesis of statistical homogeneity, truncation of the perturbative series expansion to the lowest non-zero order (first order) leads to the so-called Bourret or Born approximation \cite{Sobczyk,B1, Calvet,Turner, Bourret, Schi_SCBorn},
\begin{multline}
\Sigma^B_{k \alpha}(\textbf{q},\omega)=\hat{L}_{1 k \alpha i j}G_{i j}^0(\textbf{q},\omega)=\\ \int \ d^3s \  q_{\beta}q_{l} s_{\delta} s_{\gamma} \tilde{R}_{\gamma \alpha j l \beta k i \delta}(\textbf{q}-\textbf{s})G_{i j}^0(\textbf{s},\omega).\label{Sigma_Bo_F}
\end{multline}
Summation over repeated indices is assumed. The integral is extended to  $\mathbb{R}^3$.
The function $\tilde{R}_{\gamma \alpha j l \beta k i \delta}(\textbf{q})$ is the Fourier transform of the correlation of the elastic tensor fluctuations, $R_{\gamma \alpha j l \beta k i \delta}(\textbf{r}=\textbf{r}_1-\textbf{r}_2)=<\delta C_{\gamma \alpha j l}(\textbf{r}_1)\delta C_{\beta k i \delta}(\textbf{r}_2)>$, where $\delta \textbf{C}$ states for a fluctuation of the elastic tensor with respect to its average value $\overline{\textbf{C}}$. The self-energy in the Fourier space can thus be written as a convolution between the bare Green's dyadic and the Fourier
transform of the space-correlation of the elastic tensor fluctuations. Despite simplicity, the Born approximation imposes rather strong restrictions both on the intensity of the elastic constants fluctuations per density, $\epsilon^2$, and on the value of $\omega$ and $q$, which need to be small with respect to $a^{-1}$ \cite{B1}.
Most of the phenomenology observed in real systems, including the Rayleigh anomalies, can be qualitatively grasped even by the Born approximation \cite{Calvet}. In a vectorial framework it can even account for some features of acoustic excitations related to the coupling between longitudinal and transverse dynamics \cite{Calvet}. However, to rely on this approximation didn't permit us to unravel the presence of a well-defined shoulder in $S_L(\textbf{q},\omega)$ related to the coupling of longitudinal with transverse acoustic dynamics, as instead attested by experimental observation. This can be attributed to the limited range of validity in the wavevector space of the Born Approximation, which furthermore shifts to lower values of $q$ for larger value of $\epsilon^2$, what it is expected in the case of glasses. When $q \sim a^{-1}$ the necessary condition of validity of the Born approximation can thus be violated.
On the ground of these observations, we choose to take into account the next order of approximation in the perturbative series expansion \cite{B1}, thus obtaining the following expression for the self-energy,
\begin{equation}
\Sigma_{k \alpha}(\textbf{q},\omega)=\hat{L}_{1 k \alpha i j}<G_{ij}(\textbf{q},\omega) >^1, \label{Sigma_self1}
\end{equation}
where $<\textbf{G}(\textbf{q},\omega)>^1$ is the averaged Green dyadic calculated in the Born approximation.
On a physics ground the inclusion of second-order terms permits to take into account for multiple scattering events not covered by the Born approximation \cite{Izzo}. Second order term of the perturbative series are in the present approach accounted in an approximate form. Details are reported in the next paragraph. We thus introduce a novel method to derive an approximate expression for $\boldsymbol{\Sigma}(\textbf{q},\omega)$, which states corrective terms to the Born approximation in the framework of the perturbative series expansion. We refer to it as to a Generalized Born Approximation (GBA) \cite{Izzo}. The mathematical coherence of the GBA and its validity at wavevectors of the order of $a^{-1}$ is discussed in Ref. \cite{Izzo}. Generalizations of the Born approximation have attracted some interest \cite{Jin, Li, Zuniga}. In particular, depolarization effects in the scattering of electromagnetic waves by an isotropic random medium has been predicted by exploiting a second order representation (with respect to terms of a Neumann iteration series) for the scattered intensity \cite{Zuniga}.

Before giving the details of the expression of the self-energy stated by the GBA, we observe that the Rayleigh anomalies in glasses can be modelled by exploiting the  so-called Self-Consistent Born Approximation (SCBA) \cite{Schirmacher_4th,Kohler,Shir_gen1, Marruzzo, Ferrante}. Instead of looking to a suitable expression for the mass-operator by truncating the Neumann-Liouville series, the stochastic equation describing the disordered system is replaced by an effective nonlinear deterministic equation. The SCBA or Kraichnan model \cite{Kraichnan} reads as:
\begin{equation}
\Sigma_{k \alpha}(\textbf{q},\omega)=\hat{L}_{1 k \alpha i j}<G_{ij}(\textbf{q},\omega) >. \label{Sigma_self}
\end{equation}
 Eq. \ref{Sigma_self} together with Eq. \ref{Dyson} corresponds to successive self-consistent approximations for $\boldsymbol{\Sigma}(\textbf{q},\omega)$ and $<\textbf{G}(\textbf{q},\omega)>$. Because the effective deterministic equation can be related to a realizable model, this approximate solution guarantees certain consistency properties \cite{Kraichnan}.
At the zero-th step it is  $<\textbf{G}(\textbf{q},\omega)>=\textbf{G}^0(\textbf{q},\omega)$. The first-step of the iteration procedure thus corresponds to the Born approximation, whereas Eq. \ref{Sigma_self1} corresponds to the second-step. 
Analytical expressions in the Rayleigh region ($aq \ll 1$) can be obtained by assuming $q=0$ in the expression of $\boldsymbol{\Sigma}(\textbf{q},\omega)$ at each step of the self-consistent procedure \cite{Schirmacher_4th,Marruzzo,Ferrante}. This approach allowed a description of the Rayleigh anomalies in glasses \cite{Marruzzo, Ferrante, Schirmacher_4th} . Outcomes from this procedure are discussed in Supplementary Note 3. Even though one could expect the vectorial SCBA to carry information also about polarization properties, they can be hidden by the unfeasibility to obtain an analytical calculation at $q \sim a^{-1}$, where the wavevector dependence of the self-energy cannot be ignored and where the mixing of polarizations is expected. In terms of the perturbative series expansion the SCBA includes some terms not accounted by the Born approximation or the GBA. We, however, observe that since the convergence properties of the perturbative series expansion are unknown to consider a larger number of terms not necessarily improves the approximation.

We adopt the hypothesis of local isotropy and introduce the orthonormal basis defined by the direction of wave propagation (longitudinal) and the two orthogonal (transverse) ones  \cite{Turner}. On this basis, the \lq bare', the averaged Green's dyadic and the self-energy are diagonal. The final expression exploited to describe the diagonal element of the mass-operator in the GBA is
\begin{multline}
\Sigma_{kk}(\textbf{q},\omega) = \hat{L}_{1  k  k ii}\lim_{\eta \to 0^+}\frac{1}{\tilde{c_{i}}^2}\Big\{\ \frac{1}{\tilde{q}_{0i,\eta}^2-q^2}+ \frac{1}{[\tilde{q}_{0i,\eta}^2-q^2]^2} \cdot \\ \cdot \frac{\epsilon^2}{\tilde{c}_{i}^2}q^2\Delta\tilde{\Sigma}_{ii}^{1}(0,\omega_{\eta})\Big\},
\label{Sigma_K_ap0}
\end{multline} 
where $\tilde{q}_{0i,\eta}=\frac{\omega_{\eta}}{\tilde{c}_{i}}$,  $\omega_{\eta}=\omega+i \eta$, $\eta$ is a positive real variable, $\Delta \tilde{\Sigma}_{ii}^1(\textbf{q},\omega)=\tilde{\Sigma}_{ii}^1(\textbf{q},\omega)-\tilde{\Sigma}_{ii}^1(0, 0)$, $\tilde{\Sigma}_{ii}^1(\textbf{q},\omega)=(\epsilon^2q^2)^{-1}\hat{L}_{1 iijj}G_{jj}^0(\textbf{q},\omega)$, $\tilde{c}_i=[(c_i^0)^2+\epsilon^2\tilde{\Sigma}_{ii}^1(0, 0)]^{1/2}$, $c_i^0$ is the phase velocity of the \lq bare' medium for polarization $i=L,T$. In the next paragraph we describe the passages leading from Eq. \ref{Sigma_self1} to its approximate expression, Eq. \ref{Sigma_K_ap0}. Specifics on $\Sigma_{kk}(\textbf{q},\omega)$ computations are outlined in Appendix A.
The repeated indexes $kk, ii, jj=L,T$ with \lq L' and \lq T' labelling longitudinal and transverse directions, respectively. It is thus clear that the longitudinal and transverse self-energy are both composed by two terms, each one accounting respectively for the coupling with longitudinal and transverse dynamics, i.e. $\Sigma_{L(T)} =\Sigma_{LL(TT)}+\Sigma_{LT(TL)}$. 
We consider only spatial fluctuations of shear modulus in agreement with previous literature studies \cite{Marruzzo, Schirmacher_4th, Schi_SCBorn}. The introduction of fluctuations of the Lam\'e parameter \cite{Ferrante, Calvet, Turner} is discussed in Supplementary Note 2.
Under the hypothesis of local isotropy, $\tilde{\textbf{R}}(\textbf{q})$ can be factorized in the product of a scalar covariance function, $\tilde{r}(\textbf{q})$, and a tensor, which depends only on the angle between the versors $\hat{q}$ and $\hat{s}$ \cite{Turner}. 
We consider the simplest form of the shear modulus fluctuations scalar correlation function, in real space an exponential decay with correlation length $a$. It has been shown that this function can coherently describe the correlation function related to spatial heterogeneity in disordered systems \cite{Torquato}. In the vectorial Fourier space it corresponds to the Lorentz function,
\begin{equation}
\tilde{r}(\textbf{q})=\epsilon^2 \frac{1}{\pi^2}\frac{q^2a^{-1}}{(q^2+a^{-2})^2}. \label{R_mu_FT_exp}
\end{equation}
The integrations over the wavevector and the angular coordinates in Eq. \ref{Sigma_K_ap0} are performed analytically and numerically, respectively. The input parameters of the GBA are $a$, the disorder parameter $\epsilon^2$, and $c^0_{L(T)}$.

The hypothesis of local isotropy, finally, deserves further comments. Pure transverse modes do not contribute to the measured IXS signal in the first Brillouin zone. In solids with local anisotropy quasi-longitudinal or quasi-transverse modes (i.e. modes composed by different polarizations with a dominant polarization component) exist in the mesoscopic region. They can contribute to the measured IXS signal giving rise to the shoulder at frequencies near the characteristic frequency of the transverse excitation. It is possible, in the frame of RMT, to account for both local and statistical anisotropy \cite{Turner2}. We, however, neglected such effects here because our aim is to point out how, even in an isotropic medium, the disorder generates the mixing of polarizations.
\paragraph{Developing the Generalized Born Approximation.}
The GBA states for an approximate expression of the mass operator which includes corrective terms to the Born approximation in the context of the perturbative series expansion \cite{Izzo}.
The Born Approximation corresponds to truncate the perturbative series expansion to the first order. The next order approximation is stated in Eq. \ref{Sigma_self1}, where the stochastic operator $\hat{\textbf{L}}_1$ acts on $ <G_{ii}(\textbf{q},\omega)>^1=\lim_{\eta \rightarrow 0^+} \frac{1}{\tilde{c}_{i}^2} \big\{\frac{1}{\tilde{q}_{0i, \eta}^2-q^2-q^2\frac{\epsilon^2}{\tilde{c}_{i}^2}\Delta\tilde{\Sigma}_{ii}^1(\textbf{q},\omega_{\eta})} \big\}$. The expression in curly brackets is formally expanded in the Taylor series $\lim_{\eta \rightarrow 0^+} \sum_{n=0}^{\infty} \frac{\big[\frac{\epsilon^2}{\tilde{c}_{i}^2}q^2\Delta \tilde{\Sigma}_{ii}^{1}(\textbf{q},\omega_{\eta})\big]^n}{[\tilde{q}_{0 i, \eta}^2-q^2]^{n+1}}$. This series is convergent almost everywhere (a.e.) in a domain of the wavevector-frequency plane where the conditions $\frac{\epsilon^2}{\tilde{c}_{i}^2}|\Delta \tilde{\Sigma}^1_{ii}(\textbf{q},\omega)| < 1$ and $\mathrm{Im}[\Delta \tilde{\Sigma}^1_{ii}(\textbf{q},\omega)]>0$ are  fulfilled \cite{Izzo}. In the case that the scalar covariance function is an exponential decay it is possible to find for each polarization a bounded region of the wavevector-frequency plane, $\Omega^i=(0,\omega_{Max}) \times (0,q_{Max}^i)$, where the conditions above are fullfilled and the Taylor series is thus a.e. convergent \cite{Izzo}.  It follows that $\Sigma_{kk}(\textbf{q},\omega)$ in Eq. \ref{Sigma_self1} can be approximated by the series in the following,
\begin{multline}
\Sigma_{kk}(\textbf{q},\omega) \sim \lim_{\eta \rightarrow 0^+}  \frac{1}{\tilde{c}_{i}^2}\sum_{n=0}^{\infty}\hat{\textbf{L}}_{1kk ii}\Big\{\frac{\big[\frac{\epsilon^2}{\tilde{c}_{i}^2}q^2\Delta \tilde{\Sigma}_{ii}^{1}(\textbf{q},\omega_{\eta})\big]^{n}}{[\tilde{q}_{0i,\eta}^2-q^2]^{n+1}}\cdot \\ \cdot \theta(q_{Max}^{i}-q)\Big\}, \label{approssi}
\end{multline}
for frequencies inside and wavevectors well inside the smaller of the $\Omega^i$ domains \cite{Izzo}. The function $\theta(q)$ is the Heaviside function. The domain of the wavevector-frequency plane where Eq. \ref{approssi} holds includes both the Rayleigh region and the wavevectors region $aq \sim 1$, where the mixing of polarizations is expected \cite{Izzo}.
We truncate the Taylor series in Eq. \ref{approssi} to the first order. Furthermore, up to  wavevectors  relevant to our interest, $q = O(a^{-1})$, assuming $\Delta\tilde{\Sigma}^{1}(\textbf{q},\omega) \approx \Delta\tilde{\Sigma}^1(0,\omega)$ and then extending to infinty the upper integration boundary does not significatively enlarge the error related to the given approximation, while facilitating the analytical calculation \cite{Izzo}. Eq. \ref{Sigma_K_ap0} is thus retrieved. 

\paragraph{Analysis of theoretical outputs.}
The quantities characterizing $S_{L(T)}(q,\omega)$ calculated by GBA, i.e. broadening, phase velocity and intensity related to the inelastic features, can be derived  by a so-called spectral function approach \cite{Calvet, Sheng}, i.e by fitting the calculated dynamic structure factors with a fitting model composed by one or two DHO functions, following the same two protocols used for the IXS data. Since most of the experimental data presented in literature have been analysed with a fitting model composed by one or two DHO profiles \cite{Ruzi, Scopigno1HFglasses_transverse, zanatta_INS_GeO2, Cimatoribus, Benci, hydro_MonGio, Ruffle2}, the spectral function approach permits a connection between our theoretical outputs and literature results. 

It has been shown that the short-range structure of glasses preserves a residual order that characterizes the long-range structure of crystals \cite{Chumakov1, Giordano} with the consequent occurrence of a pseudo-Brillouin zone and of a related bending of the dispersion of the characteristic energies of acoustic excitations. We superimpose to the dynamic structure factors calculated by GBA such a bending effect (see Fig. \ref{MODEL_FIT_disp2_ord2}). This is obtained by a suitable normalization \cite{Bosak} of the frequency  of $G^0(\textbf{q},\omega)$ entering in Eq. \ref{Dyson}, i.e. $\tilde{\omega}=\omega[\frac{q_0}{q\pi} \cdot sin(\frac{\pi q}{q_0})]^{-1}$, where $\tilde{\omega}$ states for a normalized frequency and $\omega$ for unrenormalized. In order to not overload the notation in the following the tilde will be omitted. The edge of the first pseudo-Brillouin zone, $q_0=17.9$ $nm^{-1}$, is derived by the experimental dispersion of the characteristic energy of longitudinal excitations obtained by 1-DHO fitting model (see Fig. \ref{dispersion1DHO}). 
The possible broadening of the acoustic excitations due to the existence of a distribution of nearest-neighbor values is also taken into account, similarly to what is done in Ref. \cite{Giordano}. The detailed procedure is described in Supplementary Note 1. 
From $S_{L(T)}(q,E)$ obtained by GBA $g(E) / E^2$ is derived by a numerical integration,
\begin{equation}
\frac{g(E)}{E^2}=\frac{2}{\pi q_D^3}\int_0^{q_D} dq [S_L(q,E)+2S_T(q,E)],
\end{equation}
where $q_D$ is the Debye wavevector. The value of $q_D$ is obtained by the relationship $q_D=(6 \pi^2)^{\frac{1}{3}}\frac{q_0}{2 \pi}$.
\section{Discussion} \label{discussion}
\begin{figure}
\centering
\includegraphics[width=1\linewidth]{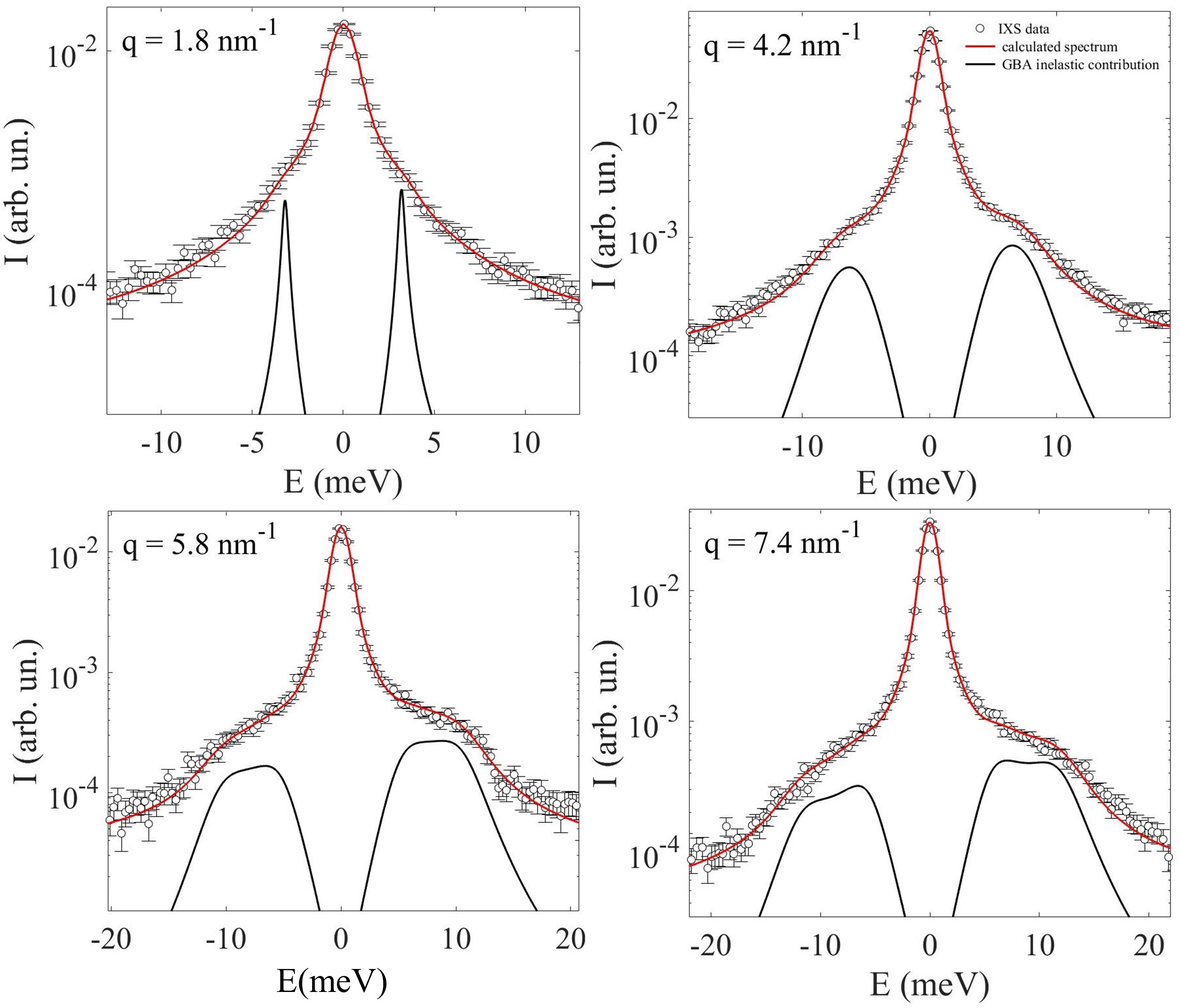}
\caption{Comparison between IXS spectra of 1-Octyl-3-methylimidazolium chloride glass for selected $q$-values (black circles with error bars) and calculated spectra (red line) obtained from GBA-modelled inelastic contribution (black line).} \label{Fig_fitGBA}
\end{figure}

\begin{figure}
\centering
\includegraphics[width=0.95\linewidth]{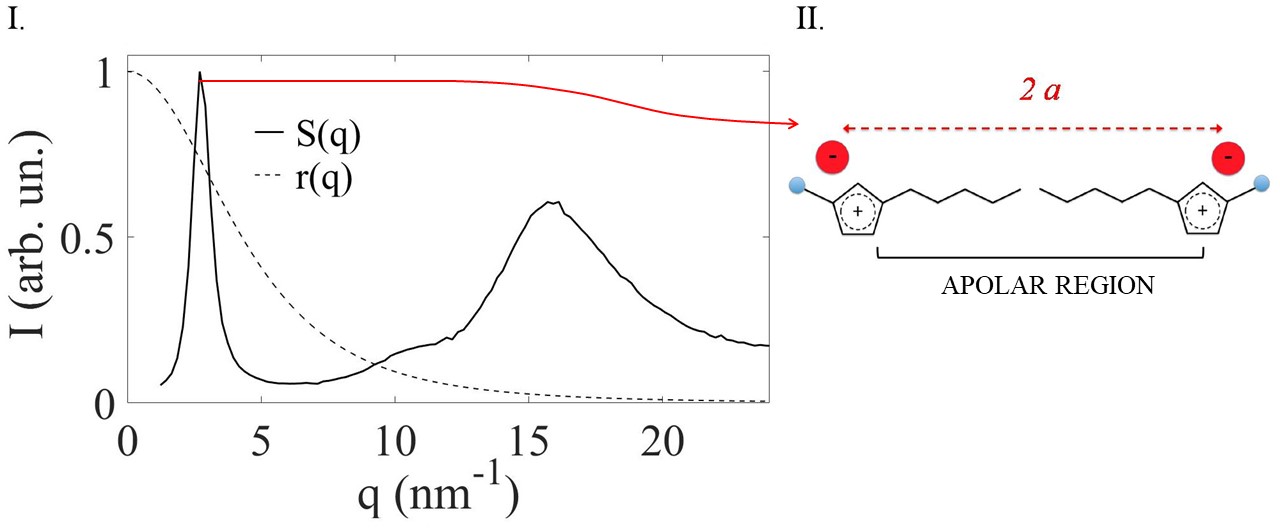}
\caption{\textit{Panel I.} Measured $S(q)$ (black line) is compared with the scalar shear modulus fluctuations correlation function (dashed line), $\tilde{r}(\textbf{q})$, used in GBA. The correlation length, $a$, is set to $\frac{\pi}{q_{FSDP}}$. 
\textit{Panel II.} Pictorial representation of the spatial configuration of the 1-octyl-methylimadozolium chloride glass at the nanoscopic scale. The aggregation of the alkyl chain generates the non-polar domains whose average size is $2 a \sim 2\pi / q_{FSDP}$.} \label{fig_Sq}
\end{figure}

\begin{figure}
\centering
\includegraphics[width=0.98\linewidth]{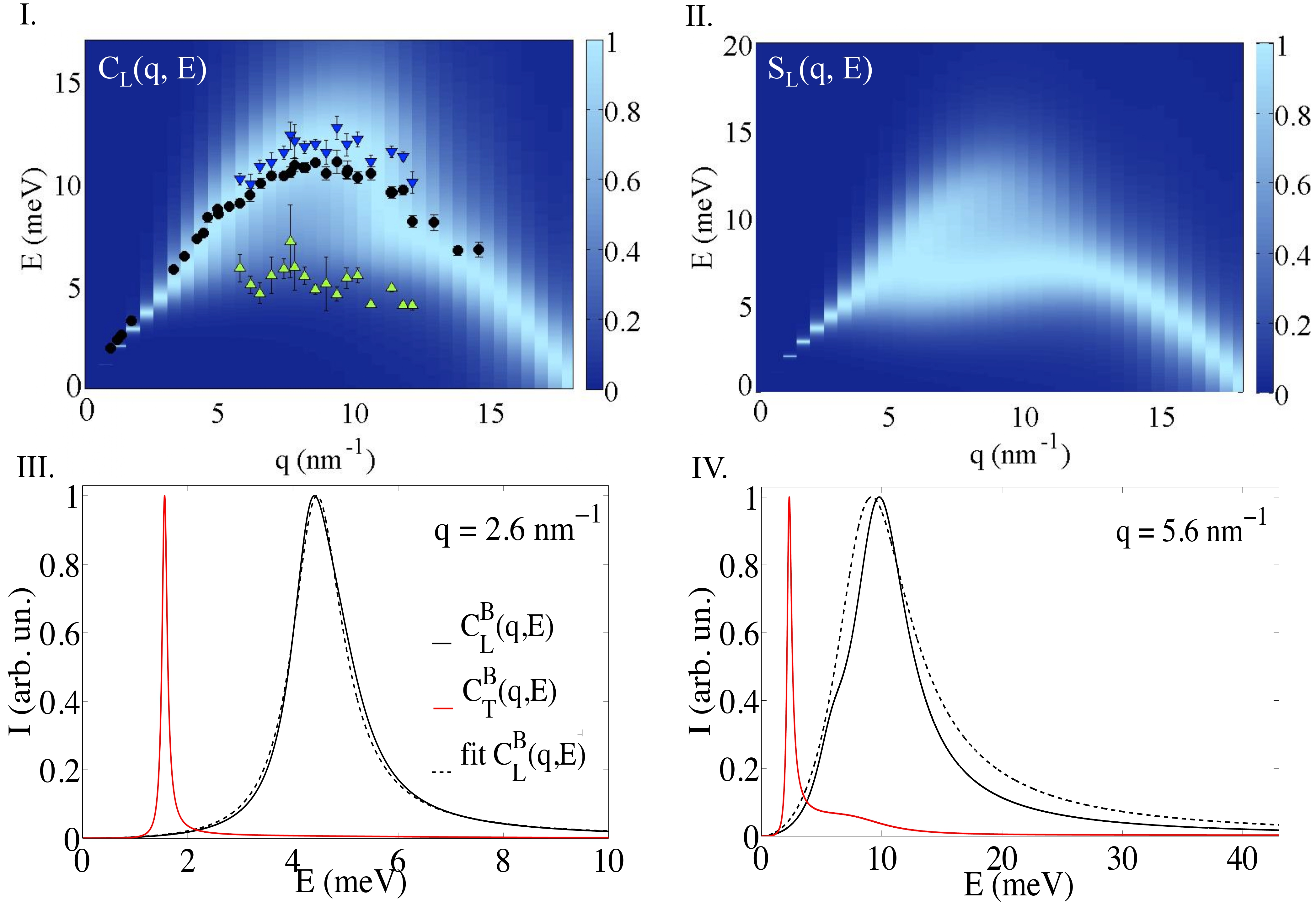}
\caption{\textit{Panel I.} Projection on the $(q,E)$ plane of $C_L(q,E)$, obtained by GBA. The frequency normalization of the \lq bare' Green dyadic in Eq. \ref{Dyson} is enforced to account for the bending of the dispersion of  the characteristic energies of acoustic excitations. 
The maximum of $C_L(q,E)$ for each $q$ is normalized to one. Black circles show the experimental values of $\Omega_L$ obtained with the 1-DHO fitting model, blue and green triangles the values obtained with the 2-DHO fitting model. \textit{Panel II.} Projection on the $(q,E)$ plane of $S_L(q,E)$. The low-energy feature related to the mixing of longitudinal and transverse polarization is here highlighted.
\textit{Panel III-IV.} Representative $C_L(q, E)$ (black line) and $C_T(q, E)$ (red line) obtained by exploiting the GBA in the low-q and high-$q$ region. The low-energy shoulder in $C_L(q,E)$, generated by the mixing of polarizations, is clearly observable. The dashed line shows the best-fit curve of $C_L(q,E)$ obtained with 1-DHO fitting model.} \label{MODEL_FIT_disp2_ord2}
\end{figure}

The comparison between IXS spectra for selected $q$ values and calculated spectra obtained by GBA is shown in Fig. \ref{Fig_fitGBA}. Before to be compared with the experimental data the theoretical $S_L(\textbf{q},\omega)$ achieved by GBA have be multiplied for the Bose factor, added to a zero-centred delta function whose intensity is an adjustable parameter and then convoluted with the measured experimental resolution. A constant adjustable background is also included. The agreement between experimental and theoretical spectra is very good. As discussed in Sec. \ref{exp_results} the quantities characterizing the longitudinal acoustic dynamics of [C8MIM]Cl glass, i.e. characteristic energies, broadening and (relative) intensities of the features of the inelastic component of the spectra, can be obtained by fitting both the experimental and theoretical spectra with an empirical model function containing one or two DHO functions. The adjustable parameters of the GBA (see Sec. \ref{GBA}) $\epsilon^2$ and $c^0_{L(T)}$ have been set by making the dispersion of the theoretical characteristic quantities of the longitudinal dynamics overlapping the experimental ones. The numerical integration on the angular coordinate can indeed overly slow down the time of convergence of a direct fit of the experimental data with a fitting model function directly obtained by GBA. The input parameter $a$ of GBA is instead fixed to $\frac{\pi}{q_{FSDP}}$. Fig. \ref{fig_Sq} shows the measured $S(q)$ and $\tilde{r}(q)$ with the assigned value of $a$. The underlying relationship between $q_{FSDP}$ and $a$ is highlighted. It can be observed in Fig. \ref{Fig_fitGBA} that for $q$ sufficiently large the GBA correctly describes the extra feature in the inelastic component observed in the IXS spectra (see Sec. \ref{exp_results}). As detailed below the comparison with GBA results permits to unambiguously  attribute such a feature to the mixing of polarizations. 
The $q$ trend of the extra feature observed in the GBA computation of $S_L(q,E)$ can be furthermore evaluated by inspection of Fig. \ref{MODEL_FIT_disp2_ord2}. It shows the projection on the $(q,E)$ plane of the dynamic structure factors and of the current spectra, $C_L(q,E) \propto \frac{E^2}{q^2} S_L(q,E)$, obtained by GBA. Current spectra for two significant $q$ values are furthermore shown. The extra feature in $S_L(q,E)$ appears at $q$ high enough, while merging in the main peak when $q$ is lowered. It is closed to the characteristic frequency of the transverse excitation, see also Ref. \cite{Izzo2}. The endorsement of the fact that such a secondary peak originates from the mixing of polarizations stems, furthermore, from the observation that it disappears when in the longitudinal self-energy, $\Sigma_L=\Sigma_{LL}+\Sigma_{LT}$, the cross term accounting for the coupling with transverse dynamics, $\Sigma_{LT}(q,\omega)$, is set to zero \cite{Izzo}. 

\begin{figure}
\centering
\includegraphics[width=1.0\linewidth]{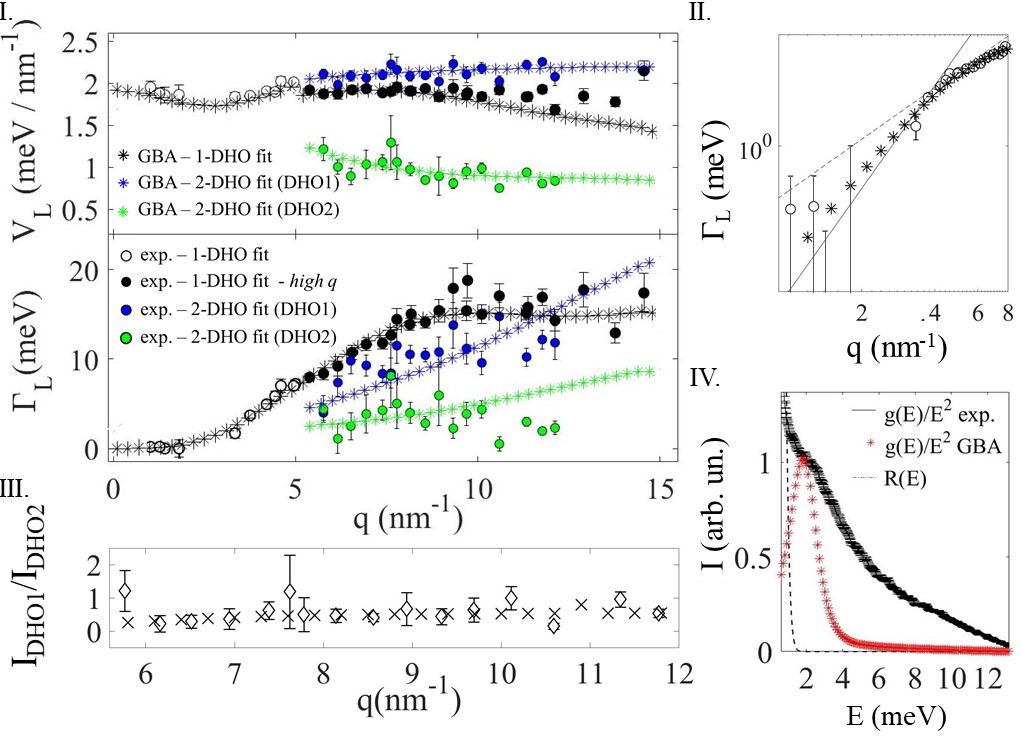}
\caption{\textit{Panel I.} Phase velocity (upper graph) and broadening (lower graph) of longitudinal acoustic excitations obtained by $IXS$ and GBA as a function of $q$. Open and black circles show the values obtained from experimental data by 1-DHO fitting model respectively in the low- and  high-$q$ region. Blue and green circles represent the values obtained by 2-DHO fitting model. Stars mark the values derived from GBA. Black stars are obtained by exploiting 1-DHO fitting model, blue and green stars 2-DHO fitting model. \textit{Panel II.}  Experimental and theoretical broadening in the low-$q$ region. Rayleigh anomaly is highlighted. Full and dashed lines are guide to eyes showing respectively a $q^4$ and $q^2$ trend. \textit{Panel III.}  Ratio of intensities of the two DHO functions modelling the inelastic features observed in the high-$q$ region observed in experimental data (diamonds) and GBA outcomes (crosses). \textit{Panel IV.} Experimental (black lines) and GBA modelled (red stars) $g(E)/E^2$. Dashed line represents the experimental energy resolution.} \label{MODEL_disp2_ord2}
\end{figure}

A detailed comparison between the experimentally and theoretically derived characteristic quantities of the longitudinal acoustic dynamics of [C8MIM]Cl glass is reported in Fig. \ref{MODEL_disp2_ord2}. 
The phase velocity, corrected for the bending of the characteristic frequency dispersion, $V_{L(T)}(q)$, shown in Fig. \ref{MODEL_disp2_ord2} is obtained through the relationship $V_{L(T)}(q)=\Omega_{L(T)}(q) \frac{\pi}{q_0 \sin(q \pi/ q_0)}$.
In the Rayleigh region ($aq \ll 1$) the dynamic structure factor related to longitudinal acoustic dynamics is characterized by a well defined inelastic excitation (see Fig. \ref{Fig_fitGBA} and Fig. \ref{MODEL_FIT_disp2_ord2}, \textit{Panel III}). The Rayleigh anomalies are properly reproduced: the phase velocity of the acoustic excitation shows a gradual softening with respect to its macroscopic ($q \rightarrow 0$ limit) value while the acoustic mode attenuation strongly increases, following the typical $q^4$ trend (Fig. \ref{MODEL_disp2_ord2}, \textit{Panels I} and \textit{III}).
At the boundary of the Rayleigh region ($aq \lesssim 1$) the phase velocity starts to increase and consequently a local minimum is observed in its trend.  Meanwhile a crossover from $q^4$ to $q^2$ power law is observed in the attenuation trend. 
Just below the transition to a $q^2$ trend, we observe a hump in the  $\Gamma$ dispersion and a rapid increase of the phase velocity. These features are related to the coupling between longitudinal and transverse polarizations since they are also strongly affected by the removing of the cross term $\Sigma_{LT}$ in $\Sigma_L$. Similar features have been reported in a theoretical characterisation of elastic excitations in polycrystalline aggregates drawn by the Born approximation and ascribed to te coupling of longitudinal with transverse acoustic dynamics \cite{Calvet}. The hump in the $\Gamma$ trend can be view as a prolongation of the $q^4$ behavior observed in the Rayleigh regime. It is stronger as bigger it is $\epsilon^2$ and as smaller $\frac{c^0_T}{c^0_L}$. This latter fact emphasizes how in the case of longitudinal acoustic dynamics at the edge of the Rayleigh region ($aq \lesssim 1$), where depolarization effects begin to affect the acoustic dynamics, the coupling of polarizations, though not manifesting in a clear peak-like feature in the dynamic structure factors, contributes to the $\Gamma$ increase. A scalar model can thus underestimate the attenuation observed in this wavevectors region. 
In the region $aq \sim 1$ the mixing of polarizations shows up, as discussed above. 
For $aq>1$ the co-existence of two excitations at different characteristic energies, finally, can be observed \cite{Calvet, B1, Baron}. This feature can be reproduced also by using a scalar Born approximation \cite{B1,Calvet}. In the case of [C8MIM]Cl glass it is observable in the transverse dynamics, as discussed in the following.

The properties of the acoustic dynamics described above, observed in the experimental characterization of the $[C8MIM]Cl$ glass and retrieved by GBA, are general for many topologically disordered systems, as confirmed by the experimental studies reported in literature, mostly achieved by IXS or INS \cite{hydro_MonGio, Ferrante, Ruzi, Scopigno1HFglasses_transverse, Giordano, zanatta_INS_GeO2}. 
A quantitative agreement between experimental findings for [C8MIM]Cl glass and GBA outcomes is obtained in the whole explored wavevectors range, as highlighted in Fig. \ref{MODEL_disp2_ord2}. 
In particular the following experimental features are quantitatively reproduced by the theory: (i) the Rayleigh anomalies; (ii) the crossover at $q_c$ from $q^4$ to $q^2$ trend in the $\Gamma_L(q)$ dispersion; (iii) the kink at $q_c$ in the $\Omega_L(q)$ dispersion; (iv) the presence of the low-energy shoulder in $S_L(q,E)$, related to the mixing of longitudinal and transverse polarization;  (v) the relative intensity of the two inelastic features, see Fig. \ref{MODEL_disp2_ord2}, \textit{Panel II}, and Fig. \ref{Fig_fitGBA} (vi) the position of the BP. We notice how in the case of [C8MIM]Cl glass the presence of the hump in the $\Gamma$-trend at the edge of the Rayleigh region can account for the shift of the crossover from $q^4$ to $q^2$ at energy slightly higher than the BP energy. The hump in the present case is enforced by the quite large value of intensity of elastic fluctuations characterizing the system, as outlined in point 2) in Sec. \ref{exp_results}.
%
\begin{figure}
\centering
\includegraphics[width=1.01\linewidth]{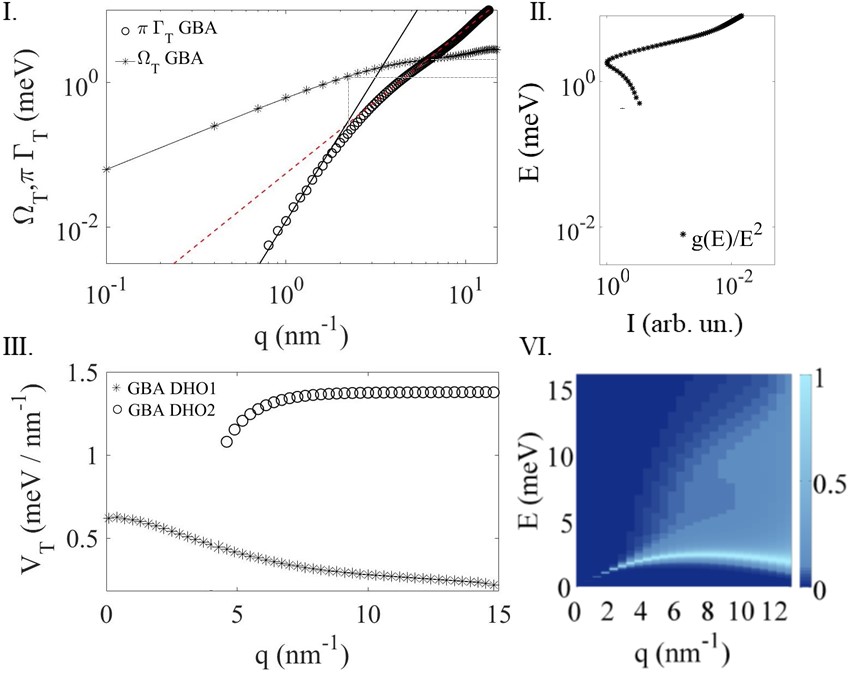}
\caption{Features of transverse dynamics obtained by GBA. \textit{Panel I.}  $\pi\Gamma_T$ (circles) and $\Omega_T$ (stars with the line) of the low-frequency inelastic feature observed in $S_T(\textbf{q},\omega)$. Full black and red dashed lines are guide to eye showing respectively $q^4$ and $q^2$ trend. The two dashed lines point out respectively the  broadening crossover from $q^4$ to $q^2$ trend and the Ioffe-Regel crossover. \textit{Panel II.} Reduced VDOS, $g(E)/E^2$. \textit{Panel III.} Phase velocities of the two inelastic excitations observed in $S_T(q,\omega)$.  \textit{Panel IV.} Projection on the $(q,E)$ plane of $C_T(q,E)$. 
} \label{MODEL_Tdisp2ord2}
\end{figure}
The input parameters of the theory used to reproduce the experimental data are $c^0_L=2.29$ $meV/nm^{-1}$, $c^0_T/c^0_L=0.52$, which is a typical value for glasses \cite{Leonforte}, $\tilde{\epsilon}^2=\frac{\epsilon^2}{\overline{\mu}^2}=0.4$, where $\overline{\mu}$ is the average shear modulus, and $a=1.0$ $nm$. The value of $a$ is set by $S(q)$ measurements. The value of $c^0_L=3.48 \ 10^3$ $m/s$ is typical for glasses, see e.g. Ref. \cite{hydro_MonGio}. There is any available measurements of speed of sound in [C8MIM]Cl glass. This value is, however, consistent with speed of sound measurements obtained for [C8MIM]Cl in liquid phase in the temperature range $T \in [280-340] \ K$ \cite{Dzida}. The value of $\tilde{\epsilon}^2$, although quite large, is meaningfully smaller than one. This is in agreement with the general trend observed in glasses, where, e.g., to fit experimental data recorded in the Rayleigh region by the SCBA it is needed to use a quite large value of the disorder parameter, near the critical value over which the system becomes unstable \cite{Ferrante}. It is furthermore in agreement with the observation reported above, which assume a quite large value of the intensity of elastic fluctuations based on the different nature of the interatomic forces acting in the two different kind of inhomogeneity domains.

Fig. \ref{MODEL_Tdisp2ord2} finally shows the features of the transverse dynamics obtained for the same input parameters used to reproduce the experimentally observed acoustic longitudinal dynamics. In the low-wavevectors region it is observed a well defined inelastic excitation characterized by: i) a crossover from $q^4$ to $q^2$ trend in the $\Gamma_T$ dispersion at the $q$-point where the characteristic frequency of the acoustic excitations matches the BP energy; ii) a softening of the phase velocity in the Rayleigh regime; iii) the Ioffe-Regel crossover at wavevector and frequency point where $\pi \Gamma_T$ becomes larger than $\Omega_T$, occurring near the BP energy. At larger $q$ a high-energy shoulder appears in $C_T(\textbf{q},\omega)$ (Fig. \ref{MODEL_FIT_disp2_ord2}). This is not entirely related to the mixing of polarizations. It is indeed partially preserved when the term $\Sigma_{TL}$ is set to zero. These results are in qualitative agreement with MD results in glasses \cite{Mossa, Marruzzo, Shintani} or liquids at high-wavevectors \cite{Ribeiro2,Bryk1}.
The characterisation of the transverse dynamics that emerges from the present study is further validated by the agreement between the characteristic energy of the BP predicted through GBA and observed by INS. The total VDOS is indeed mainly determined by transverse dynamics \cite{Marruzzo}. The VDOS behaviour for energies higher than the BP energy can be reproduced only if IVMs are taken under account, as discussed in Appendix B.
\section{Conclusion} \label{conclusion}
Within the framework of the RMT we developed a simple enough, mathematically tractable approximation allowing a unified description of the acoustic dynamics at mesoscale in glasses. The proposed model describes \textit{all} the anomalies of acoustic excitations characterizing an elastically disordered heterogeneous medium up to nanoscale, i.e. both the Rayleigh anomalies and the mixing of longitudinal and transverse polarization observed at higher wavevectors in several topologically disordered systems. The insights which emerge are in agreement with experimental observations reported in literature. The model, furthermore, is able to make quantitative and verifiable predictions, as supported by the achievement of a complete description of the acoustic dynamics in [C8MIM]Cl glass, experimentally characterized by IXS and INS in the present study. The theoretical framework built up can be thought as a starting point for describing acoustic dynamics in different kind of glasses, composites, ceramics, geophysical systems, or propagation of different kind of waves in disordered media.

The present study also emphasizes how the large tunability of the local structure of ionic liquids, which can be directed by controlling chemical and geometrical molecular structure, as well as the fact that their nanoscale heterogeneous structure has been largely attested and analyzed in literature, can be exploited to test theoretical models.
\begin{acknowledgments}
The authors thank W. Schirmacher, G. Pastore and J. Dyre for useful discussions. M. Krisch and A. Bosak are acknowledged for support during the measurements at the ID28 beamline (ESRF), J. Taylor and R. Stuart during the measurements at MARI (ISIS).
\end{acknowledgments}

\section{Appendix A}
\renewcommand\theequation{A\arabic{equation}}
\setcounter{equation}{0}
\paragraph{Specifics of $\Sigma_{L(T)}(q,\omega)$ in the Generalized Born Approximation.} 
In the orthonormal basis defined by the direction of wave propagation, $\hat{q}$, and the two orthogonal ones the average Green dyadic states
\begin{eqnarray}
<\textbf{G}(\textbf{q},\omega)>=<g_L(\textbf{q},\omega)>\hat{q}\hat{q}+<g_T(\textbf{q},\omega)>(I-\hat{q}\hat{q}). \nonumber \\ \nonumber
\end{eqnarray}
Similar expressions can be retrieved for $G^0_{L(T)}(\textbf{q},\omega)$ and $\Sigma_{L(T)}(\textbf{q},\omega)$.
%
As it follows from the properties of the inverse tensor, it is
\begin{eqnarray}
&&<g_{L(T)}(\textbf{q},\omega)>=\frac{1}{g_{L(T)}^0(\textbf{q},\omega)^{-1}-\Sigma_{L(T)}(\textbf{q},\omega)}. \nonumber
\end{eqnarray}
The bare Green's functions are
\begin{eqnarray}
&&g^0_{L(T)}(\textbf{q},\omega)=lim_{\eta \rightarrow 0^+}\frac{1}{\big(c_{L(T)}^0 \big)^2}\frac{1}{\Big(\frac{\omega+i\eta}{c_{L(T)}^0}\Big)^2-q^2}. \nonumber
\end{eqnarray}
In the GBA each partial term of the self-energy, $\Sigma_{kkii}(\textbf{q},\omega)$, is composed by two terms, see Eq. \ref{Sigma_K_ap0}, i.e.
\begin{eqnarray}
&&\Sigma_{kkii}(\textbf{q},\omega)=\Sigma_{kkii}^{(0)}(\textbf{q},\omega)+\Sigma_{kkii}^{(1)}(\textbf{q},\omega), \nonumber 
\end{eqnarray}
with
\begin{eqnarray}
\Sigma_{kkii}^{(0)}(\textbf{q},\omega)= \hat{L}_{1  k  k ii} \ lim_{\eta \rightarrow 0^+} \frac{1}{\tilde{c_{i}}^2}\frac{1}{\tilde{q}_{0i,\eta}^2-q^2}; \nonumber
\end{eqnarray}
\begin{eqnarray}
\Sigma_{kkii}^{(1)}(\textbf{q},\omega)= \hat{L}_{1  k  k ii} \ lim_{\eta \rightarrow 0^+} \frac{1}{\tilde{c_{i}}^2}\frac{\frac{\epsilon^2}{\tilde{c}_{i}^2}q^2\Delta\tilde{\Sigma}_{ii}^{1}(0,\omega_{\eta})}{[\tilde{q}_{0i,\eta}^2-q^2]^2}. \nonumber 
\end{eqnarray}
We define $\tilde{q}_{0i}=\frac{\omega}{\tilde{c}_i}$, redefine for sake of simplicity $\frac{\eta}{\tilde{c}_i}=\eta$, and use the explicit expression of the operator $\hat{\textbf{L}}_1$ stated in Eq. 2. Under the local isotropy hypothesis, in the orthonormal basis defined above, it is
\begin{widetext}
\begin{multline}
\Sigma_{LL}^{(0)}(\textbf{q},\omega)=\lim_{\eta\rightarrow 0^+}\int \hat{q}\hat{q} \hat{s}\hat{s} \ d^3s \ \tilde{r}(\textbf{q}-\textbf{s})  \frac{1}{\tilde{c}_L^2}\frac{1}{(\tilde{q}_{0L}+i\eta)^2-q^2}; \\ \Sigma_{LL}^{(1)}(\textbf{q},\omega)= \lim_{\eta\rightarrow 0^+}\int \hat{q}\hat{q} \hat{s}\hat{s} \ d^3s \ \tilde{r}(\textbf{q}-\textbf{s}) \frac{1}{\tilde{c}_L^2} \frac{\frac{\epsilon^2}{\tilde{c}_L^2}s^2\Delta \tilde{\Sigma}_L^1(0,\omega+i\eta)}{[(\tilde{q}_{0L}+i\eta)^2-q^2]^2} ; \label{ll_ii}
\end{multline}
\begin{multline}
\Sigma_{LT}^{(0)}(\textbf{q},\omega)= \lim_{\eta \rightarrow 0^+}\int \hat{q}\hat{q}(\textbf{I}-\hat{s}\hat{s}) \ d^3s \ \tilde{r}(\textbf{q}-\textbf{s}) \frac{1}{\tilde{c}_T^2} \frac{1}{(\tilde{q}_{0T}+i\eta)^2-q^2}; \\ \Sigma_{LT}^{(1)}(\textbf{q},\omega)= \lim_{\eta \rightarrow 0^+} \int \hat{q}\hat{q}(\textbf{I}-\hat{s}\hat{s}) \ d^3s \ \tilde{r}(\textbf{q}-\textbf{s}) \frac{1}{\tilde{c}_T^2} \frac{\frac{\epsilon^2}{\tilde{c}_T^2}s^2\Delta \tilde{\Sigma}_T^1(0,\omega+i\eta)}{[(\tilde{q}_{0T}+i\eta)^2-q^2]^2}; 
\label{lt_ii} 
\end{multline}
\begin{multline}
\Sigma_{TT}^{(0)}(\textbf{q},\omega)=\frac{1}{2}\lim_{\eta \rightarrow 0^+} \int (\textbf{I}-\hat{q}\hat{q}) (\textbf{I}-\hat{s}\hat{s}) \ d^3s \ \tilde{r} (\textbf{q}-\textbf{s})  \frac{1}{\tilde{c}_T^2}\frac{1}{(\tilde{q}_{0T}+i\eta)^2-q^2}; \\ \Sigma_{TT}^{(1)}(\textbf{q},\omega)= \frac{1}{2}\lim_{\eta \rightarrow 0^+} \int (\textbf{I}-\hat{q}\hat{q}) (\textbf{I}-\hat{s}\hat{s}) \ d^3s \ \tilde{r} (\textbf{q}-\textbf{s}) \frac{1}{\tilde{c}_T^2} \frac{\frac{\epsilon^2}{\tilde{c}_T^2}s^2\Delta \tilde{\Sigma}_T^1(0,\omega+i\eta)}{[(\tilde{q}_{0T}+i\eta)^2-q^2]^2};
\label{tt_ii}
\end{multline}
\begin{multline}
\Sigma_{TL}^{(0)}(\textbf{q},\omega)= \frac{1}{2}\lim_{\eta \rightarrow0^+}\int (\textbf{I}-\hat{q}\hat{q}) \hat{s}\hat{s} \ d^3s \ \tilde{r}(\textbf{q}-\textbf{s})  \frac{1}{\tilde{c}_L^2}\frac{1}{(\tilde{q}_{0L}+i\eta)^2-q^2}; \\ \Sigma_{TL}^{(1)}(\textbf{q},\omega)= \frac{1}{2} \lim_{\eta \rightarrow0^+}\int (\textbf{I}-\hat{q}\hat{q})\hat{s}\hat{s} \ d^3s \ \tilde{r}(\textbf{q}-\textbf{s})  \frac{1}{\tilde{c}_L^2} \frac{\frac{\epsilon^2}{\tilde{c}_L^2}s^2\Delta \tilde{\Sigma}_L^1(0,\omega+i\eta)}{[(\tilde{q}_{0L}+i\eta)^2-q^2]^2}.
\label{tl_ii}
\end{multline}
\end{widetext}
By performing the tensor products and using spherical coordinates \cite{Turner, Calvet} in Eqs. \ref{ll_ii}-\ref{tl_ii} we obtain
\begin{widetext}
\begin{multline}
\Sigma^{(0)}_{LL(LT)}(\textbf{q},\omega)=\epsilon^2  q^2  \int_{-1}^{+1} dx L^{\mu \mu}_{LL(LT)}(x) \frac{1}{\tilde{c}_{L(T)}^2} \frac{2}{\pi} a^{-1}  I^{(0)}_{L(T)}(\textbf{q},\omega,x); \\ \Sigma^{(1)}_{LL(LT)}(\textbf{q},\omega)=\epsilon^2  q^2  \int_{-1}^{+1} dx L^{\mu \mu}_{LL(LT)}(x) \frac{1}{\tilde{c}_{L(T)}^2}  \frac{2}{\pi} a^{-1}  I^{(1)}_{L(T)}(\textbf{q},\omega,x) \frac{\epsilon^2}{\tilde{c}_{L(T)}^2}\Delta \tilde{\Sigma}_{L(T)}^1(0,\omega);
\label{I_Sigma_LL} 
\end{multline}
\begin{multline}
\Sigma^{(0)}_{TT(TL)}(\textbf{q},\omega)=\epsilon^2  q^2  \frac{1}{2}\int_{-1}^{+1} dx L^{\mu \mu}_{TT(TL)}(x) \frac{1}{\tilde{c}_{T(L)}^2} \frac{2}{\pi} a^{-1}  I^{(0)}_{T(L)}(\textbf{q},\omega,x); \\  \Sigma^{(1)}_{TT(TL)}(\textbf{q},\omega)=\epsilon^2  q^2  \frac{1}{2} \int_{-1}^{+1} dx L^{\mu \mu}_{TT(TL)}(x) \frac{1}{\tilde{c}_{T(L)}^2} \frac{2}{\pi} a^{-1}  I^{(1)}_{T(L)}(\textbf{q},\omega,x) \frac{\epsilon^2}{\tilde{c}_{T(L)}^2}\Delta \tilde{\Sigma}_{T(L)}^1(0,\omega);
\label{I_Sigma_TT}
\end{multline}
\end{widetext}
where $x=cos(\hat{qs})$, being $\hat{qs}$ the angle between the two versors $\hat{q}$ and $\hat{s}$, and $L^{\mu \mu}_{LL}(x)=4x^4$, $L^{\mu \mu}_{LT(TL)}=4(1-x^2)x^2$, $L^{\mu \mu}_{TT}=1-3x^2+4x^4$.
By exploiting the Sokhotski-Plemelj-Fox theorem \cite{Galapon, Fox}, integration by part and the Cauchy's Residue Theorem the integral $I_k^{(n)}(\textbf{q},\omega,x)$ for a polarization $k=L,T$ and $n=0,1$, is given by 
\begin{widetext}
\begin{multline}
I_k^{(n)}(\textbf{q},\omega,x)=\lim_{\eta \rightarrow 0^+}\int_{0}^{\infty}ds \  s^2 \frac{s^2}{[a^{-2}+q^2+s^2-2qsx]^2} \frac{[s^2]^n}{[(\tilde{q}_{0k}+i\tilde{\eta})^2-s^2]^{n+1}}= \# \int_{0}^{\infty}ds \  s^2\frac{s^2}{[a^{-2}+q^2+s^2-2qsx]^2}  \cdot \\ \cdot \frac{[s^2]^n}{[(\tilde{q}_{0k}+i\tilde{\eta})^2-s^2]^{n+1}}+  (-1)^{n+1}\frac{i\pi}{n!} \frac{d^n}{dz^n}\Big\{z^2 \frac{z^2}{[a^{-2}+q^2+z^2-2qzx]^2}  \frac{[z^2]^n}{[\tilde{q}_{0k}+z]^{n+1}}\Big\}|_{z=\tilde{q}_{0k}}. \label{In}
\end{multline}
\end{widetext}
We recognize in the second term of the second side of Eq. \ref{In} the residue of order $n$ of the integrand around the pole $\tilde{q}_{0k}$. The symbol $\#$ states for the Hadamard finite part integral (equal to the Cauchy principal value when $n=0$). 
By using integration by parts, it follows that
\begin{widetext}
\begin{multline}
\# \int_{0}^{\infty}ds \ s^2 \frac{s^2}{[a^{-2}+q^2+s^2-2qsx]^2} \frac{[s^2]^n}{[\tilde{q}_0^2-s^2]^{n+1}}=\\ p.v. \int_{0}^{\infty}ds \frac{1}{n!} (-1)^{n+1}\frac{1}{(\tilde{q}_0-s)}\frac{d^n}{ds^n}\big\{s^2 \frac{s^2}{[a^{-2}+q^2+s^2-2qsx]^2} \frac{[s^2]^n}{[\tilde{q}_0+s]^{n+1}}\big\} .  \label{pv_B}
\end{multline}
\end{widetext}
The Hadamard finite part integral exists because it exists the Cauchy principal value of the integral in the second side of Eq. \ref{pv_B}, since the integrand satisfy the Lipschitz property. The Hadamard finite part integral  in Eq. \ref{pv_B} can be calculated by exploiting the Residue Theorem because the integrand has only non-essential singularities. 
\begin{figure}
\centering
\includegraphics[scale=0.195]{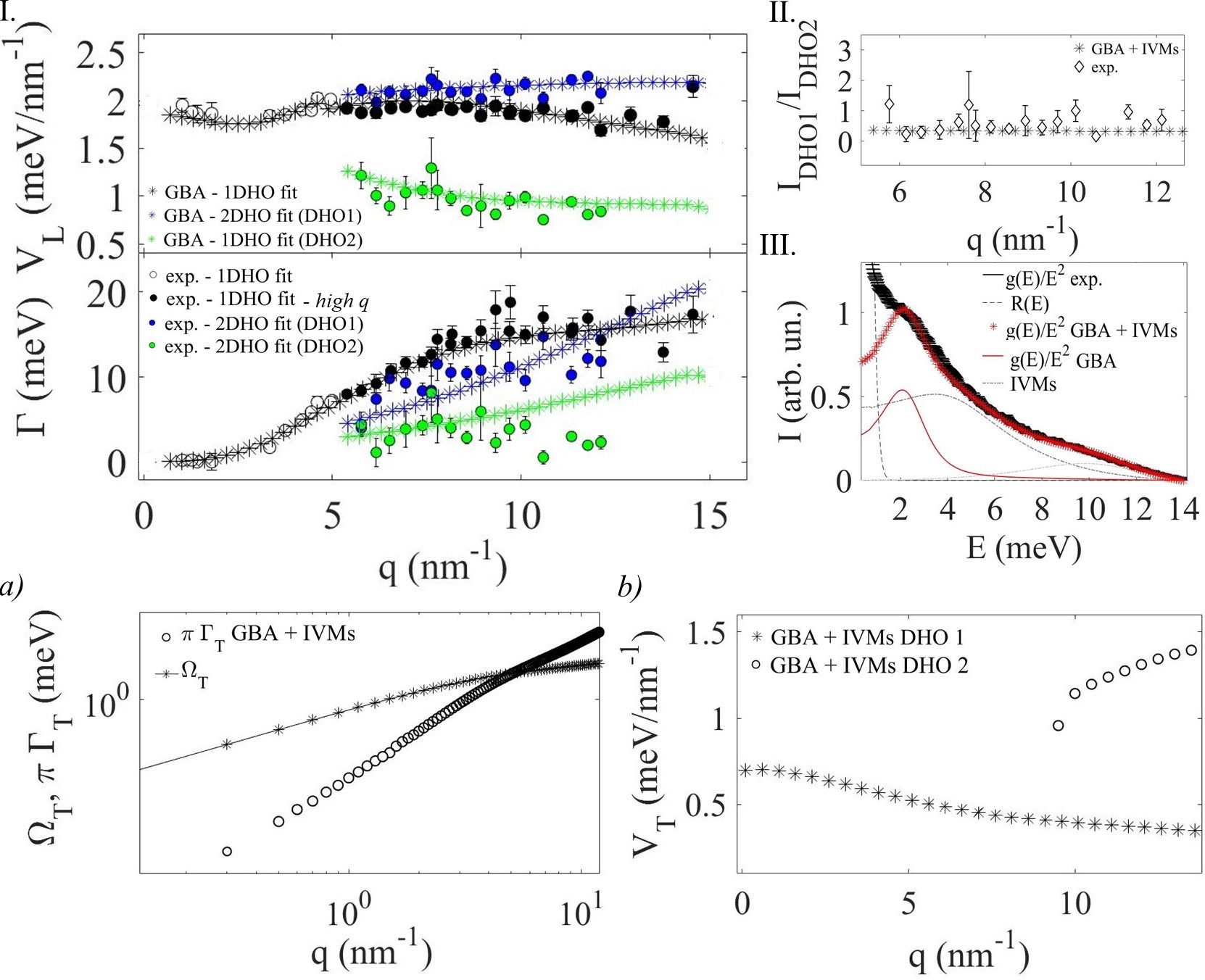} 
\caption{\textit{Longitudinal dynamics.}  \textit{Panel I.} Phase velocity (upper graph) and broadening (lower graph) as a function of $q$ of longitudinal acoustic excitations obtained by $IXS$ data (circles) and GBA derived acoustic excitations coupled to IVMs (stars). Open and black circles show  experimental outcomes obtained  by 1-DHO fitting model respectively in the low- and high-q region, blue and green circles display the values obtained by 2-DHO fitting model. Stars mark the values derived from a spectral function analysis of theoretical outcomes, black stars show the results obtained by 1-DHO fitting model, blue and green stars by 2-DHO fitting model. \textit{Panel II.}  Ratio of intensities of the DHO functions related to the two inelastic features in the high-q region. \textit{Panel III.} Broadening of inelastic excitations observed in experiment (circles with error bars) and theory (stars). \textit{Panel IV.} Experimental (black lines) and theoretical (red stars) reduced VDOS, $g(E)/E^2$. Red and dot-dashed lines show respectively the components of the reduced VDOS related to acoustic dynamics and IVMs. Dashed line show the experimental INS resolution. \textit{Transverse dynamics.} \textit{Panel a.} $\Gamma_T$, multiplied by $\pi$ (circles) and $\Omega_T$ (stars with the line), of the transverse inelastic feature of lower characteristic energy observed in $S_T(\textbf{q},\omega)$. \textit{Panel b.} Phase velocity of the two features of the inelastic components observed in $S_T(\textbf{q},\omega)$. } \label{LS}
\end{figure}
In Eqs. \ref{ll_ii}-\ref{I_Sigma_TT} it is $\Delta \tilde{\Sigma}_{L(T)}^1(\textbf{q},\omega)=\tilde{\Sigma}_{L(T)}^1(\textbf{q},\omega)-\tilde{\Sigma}_{L(T)}^1(0,0)$, where $\tilde{\Sigma}_{L(T)}$ is related to the first step self-energy obtained through the Born approximation, $\Sigma_{L(T)}^1(\textbf{q},\omega)$,
\begin{widetext}
\begin{multline}
\Sigma_L^1(\textbf{q},\omega)=\epsilon^2q^2\tilde{\Sigma}_L^1(\textbf{q},\omega)= \epsilon^2q^2[\tilde{\Sigma}_{LL}^1(\textbf{q},\omega)+\tilde{\Sigma}_{LT}^1(\textbf{q},\omega)] = \int \hat{q}\hat{q} \ \tilde{r}(\textbf{q}-\textbf{s})[g_L^0(s,\omega)\hat{s}\hat{s}+g_T^0(s,\omega)(\textbf{I}-\hat{s}\hat{s})]d^3s; \label{Sigma_L_app}
\end{multline}
\begin{multline}
\Sigma_T^1(\textbf{q},\omega)=\epsilon^2q^2\tilde{\Sigma}_T^1(\textbf{q},\omega)= \epsilon^2q^2[\tilde{\Sigma}_{TT}^1(\textbf{q},\omega)+\tilde{\Sigma}_{TL}^1(\textbf{q},\omega)] = \frac{1}{2}\int (\textbf{I}-\hat{q}\hat{q}) \tilde{r}(\textbf{q}-\textbf{s})[g_L^0(s,\omega)\hat{s}\hat{s}+g_T^0(s,\omega) (I-\hat{s}\hat{s})]d^3s. \label{Sigma_T_app}
\end{multline}
\end{widetext}
By comparing Eqs. \ref{Sigma_L_app} and \ref{Sigma_T_app} with Eqs. \ref{ll_ii}-\ref{tl_ii} it is immediate to verify that $\Sigma^1_{L(T)}$ is equivalent to $\Sigma^{(0)}_{L(T)}$ under the substitution $\tilde{c}_{L(T)} \rightarrow c_{L(T)}^0$. 
\section{Appendix B} \label{appendix}
\renewcommand\theequation{B\arabic{equation}}
\setcounter{equation}{0}
\paragraph{Coupling of acoustic excitations with Intermolecular Vibrational Modes.}
The reduced VDOS of $[C8MIM]Cl$ glass in the energy region $[7,14] \ meV$ is characterised by a broad peak-like feature, related to the presence of IVMs. They have a characteristic frequency inside this energy interval, as attested by Raman spectroscopy \cite{Ribeiro_Raman2}.
An IVM can couple to acoustic excitations, affecting their effective phase velocity and attenuation when its characteristic frequency matches the one of the acoustic excitations \cite{Cla, Duval}. 
To account for the coupling between acoustic excitations, described in the GBA framework, and IVMs we introduce an additional term, $\Sigma^{IVM}_{L(T)}$, to the self-energy calculated by GBA, $\Sigma_{L(T)}$. The coupling of the IVM with acoustic excitations is treated to lowest order \cite{Shir_gen1}. It is thus $\Sigma^{IVM}_{L(T)}=\frac{q^2A^s_{L(T)}}{\omega^2-\omega_0^2+i\omega \Gamma_s}$, where $\omega_0$ is the characteristic frequency of the IVM, $\Gamma_s$ the broadening and  $A^s_{L(T)}$ is a polarization-dependant coupling factor. The total self-energy $\Sigma^{t}_{L(T)}=\Sigma_{L(T)}+\sum_n [\Sigma^{IVM_n}_{L(T)}]$ enters into the expression of $<G_{L,T}(\textbf{q},\omega)>$ through Eq. 1 in the text. $n$ indexes IVMs. We furthermore introduce the IVM's dynamic structure factor and define the corresponding IVM's contribution to the reduced VDOS,
\begin{eqnarray}
S^{IVM}(\textbf{q},\omega)=\frac{1}{\pi}\frac{q^2}{\omega}
\textit{Im}\big\{[\omega^2-\omega_0^2+ i \omega \Gamma_s+ q^2 (A_{L}^s \nonumber \\ <g_L(\textbf{q},\omega)>+2 A_{T}^s<g_T(\textbf{q},\omega)>)]^{-1}\big\} \nonumber;
\end{eqnarray}
\begin{eqnarray}
\frac{g^{IVM}(E)}{E^2}=\frac{2}{\pi q_D^3}\int_0^{q_D} dq S^{IVM}(\textbf{q},\omega). \ \ \ \ \ \ \ \ \ \ \ 
\nonumber
\end{eqnarray} 
The contribution of IVMs to the reduced VDOS, $\sum_n \frac{g^{IVM_n}(E)}{E^2}$, is added to that related to the acoustic dynamics to obtain $\frac{g(E)}{E^2}$. 
To cope with literature data \cite{Ribeiro_Raman2} we introduce two IVMs with respectively characteristic frequency and attenuation $\omega_{0}^{(1)}=7.2$ $meV$, $\Gamma_s^{(1)}=7.5$ $meV$ and $\omega_{0}^{(2)}=11$ $meV$, $\Gamma_s^{(2)}=6.5$ $meV$. The value of the parameters $A_{L(T)}^{s(1,2)}$ have been set in order to reproduce the measured VDOS. 
Outcomes from this modelling are shown in Fig. \ref{LS} for both longitudinal and transverse dynamics. They are contrasted against the corresponding experimental features. The input parameters of the GBA are the same when IVMs are not considered, but $c^0_T/c^0_L$ is moved from $0.52$ to $0.54$. Inspection of Fig. \ref{LS} permits to infer that the use of GBA combined to the lowest order modelling of the coupling between acoustic excitations and IVMs allows to achieve a quantitative agreement with experimental outcomes.
We notice that the high-frequency region of the reduced VDOS can be described only after the introduction of IVMs.
The presence of IVMs can also influences the broadening of $S_{L(T)}(\textbf{q},\omega)$ at frequencies higher than BP frequency, as well as the position of the BP in VDOS. It does not account for the mixing of polarization of acoustic excitations, which can be instead described only by GBA.

\end{document}